%% file: PhishingOnDemand.tex
\documentclass[conference]{IEEEtran}

\input{preamble}

\begin{document}
\hyphenation{op-tical net-works semi-conduc-tor}

\newcommand\phod{our prototype\xspace}
\newcommand\Phod{Our prototype\xspace}
\newcommand\Template{\emph{Template-based} solutions\xspace}
\newcommand\Handcrafted{\emph{Handcrafted} solutions\xspace}

\newcommand\iconEye{\includegraphics[width=0.01\textwidth,trim=0cm 2.0cm 0cm 0cm]{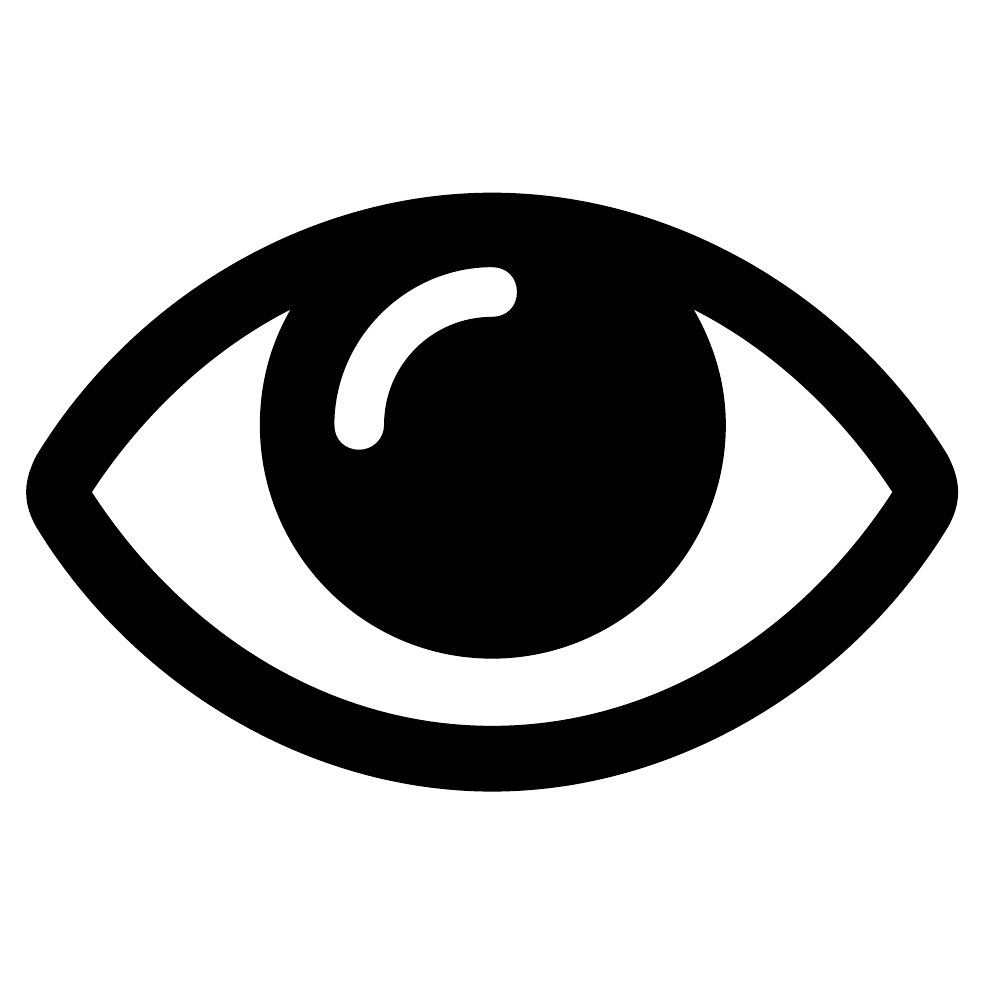}\xspace}
\newcommand\iconLightning{\includegraphics[width=0.01\textwidth,trim=0cm 2.0cm 0cm 0cm]{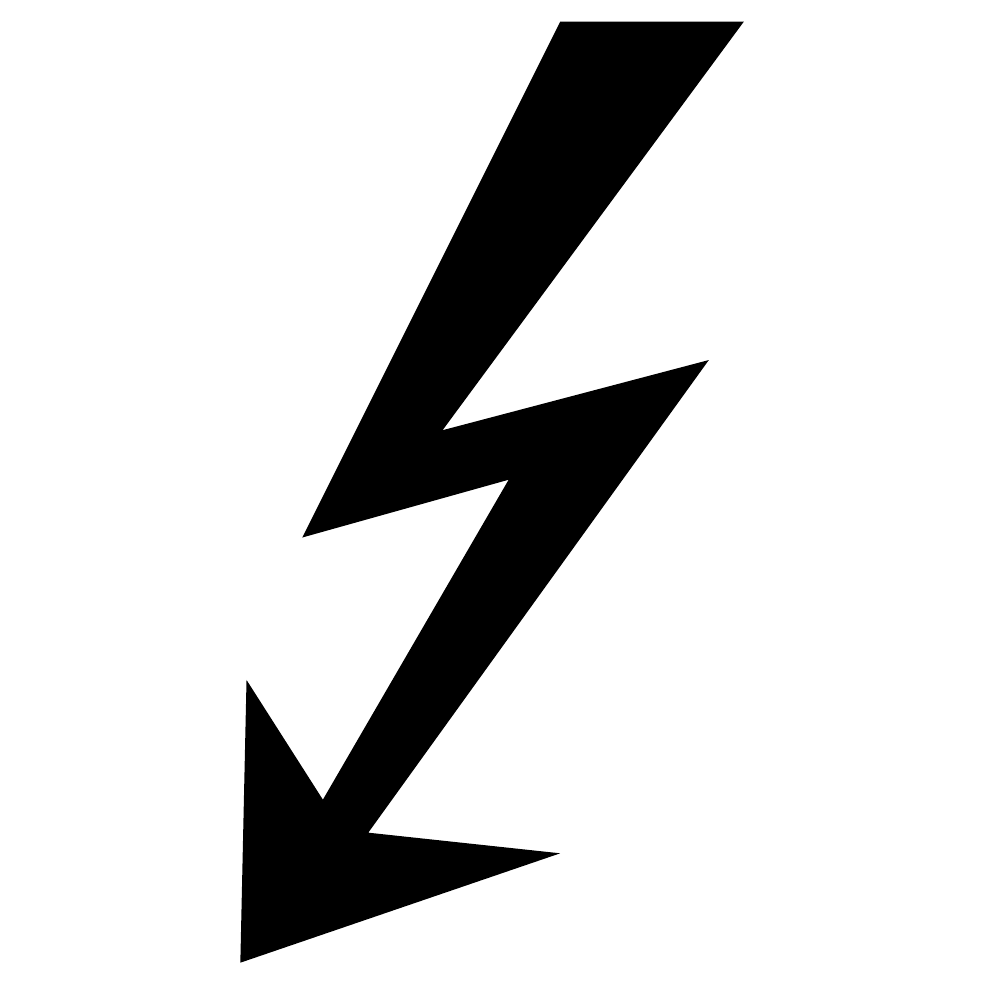}\xspace}
\newcommand\iconStar{\includegraphics[width=0.01\textwidth,trim=0cm 2.0cm 0cm 0cm]{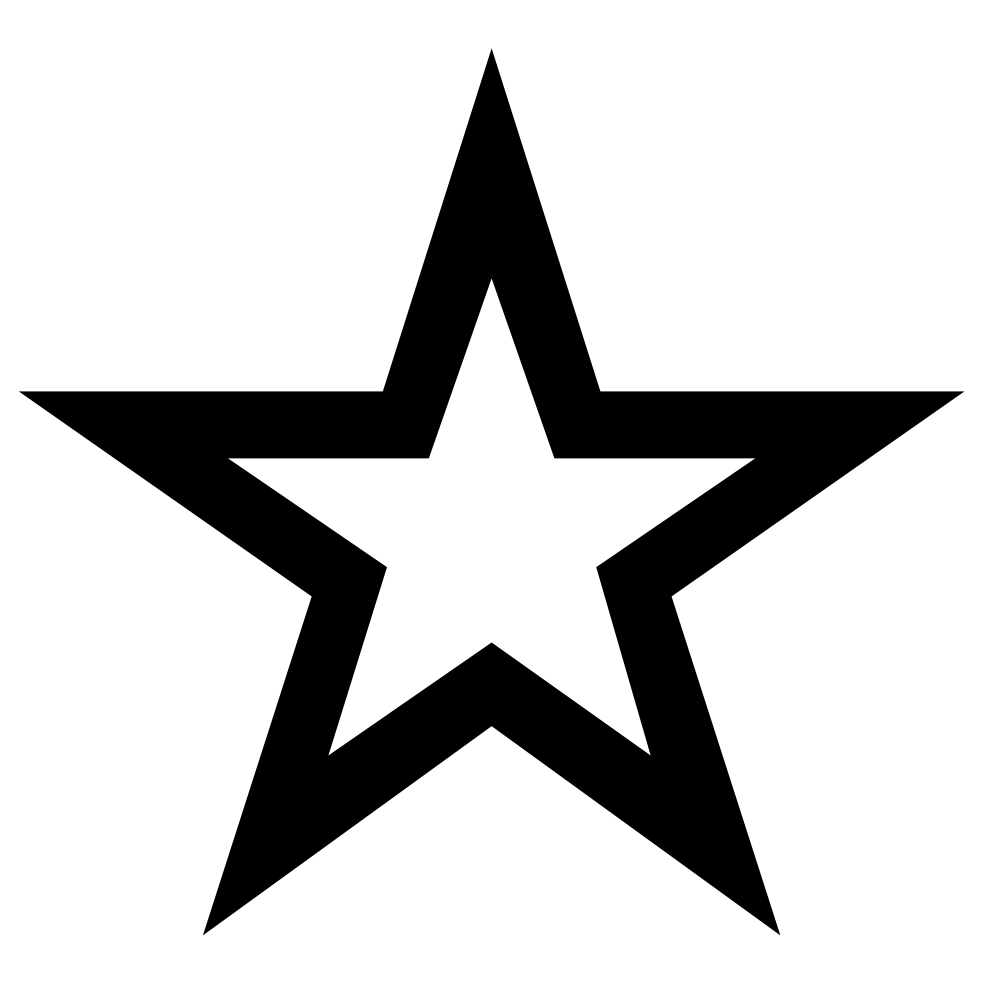}\xspace}
\newcommand\iconStop{\includegraphics[width=0.01\textwidth,trim=0cm 2.0cm 0cm 0cm]{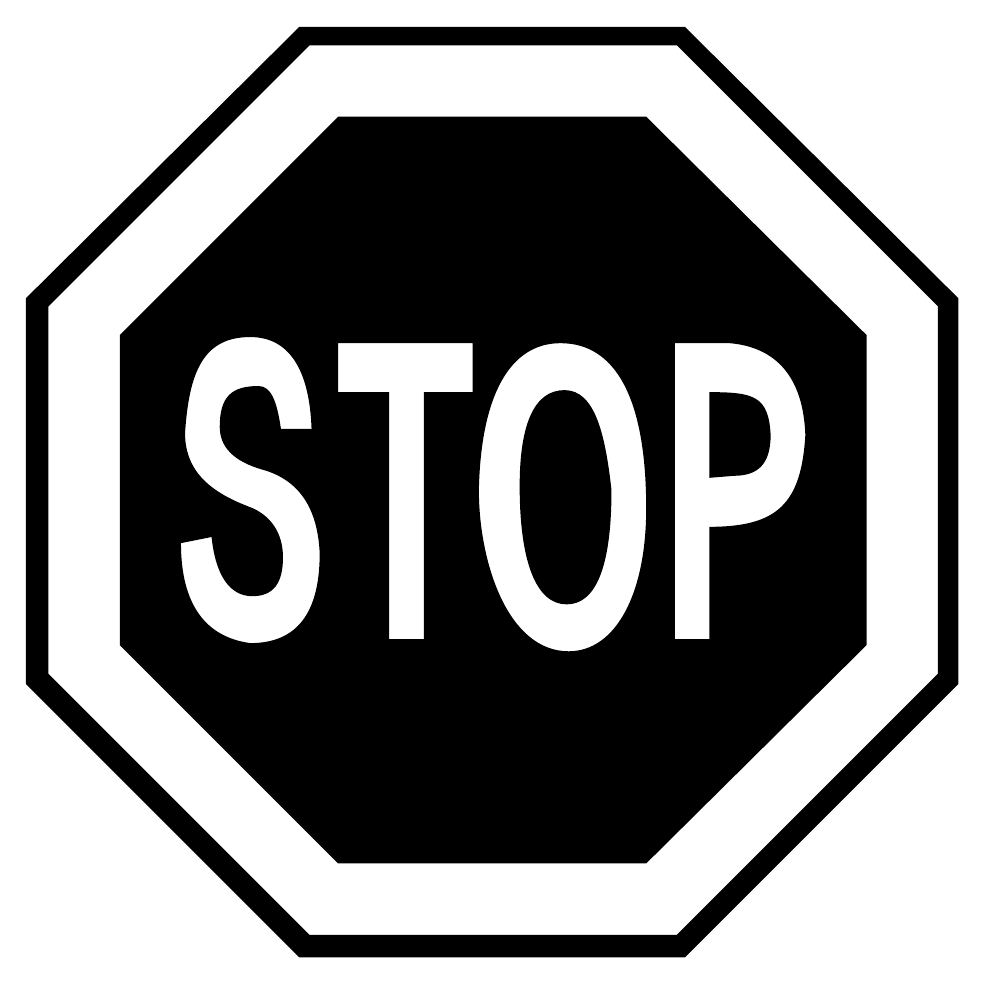}\xspace}
\newcommand\iconDisconnect{\includegraphics[width=0.01\textwidth,trim=0cm 2.0cm 0cm 0cm]{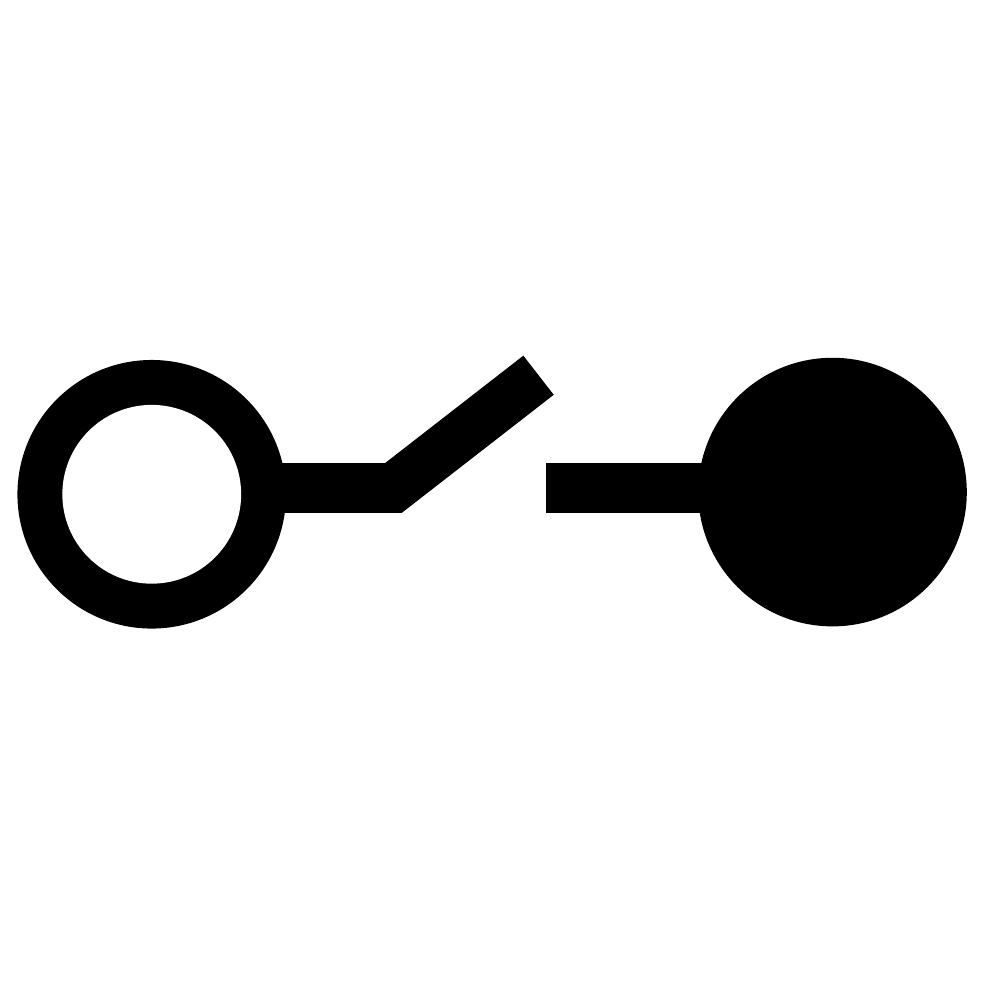}\xspace}
\newcommand\iconTemplate{\includegraphics[width=0.01\textwidth,trim=0cm 2.0cm 0cm 0cm]{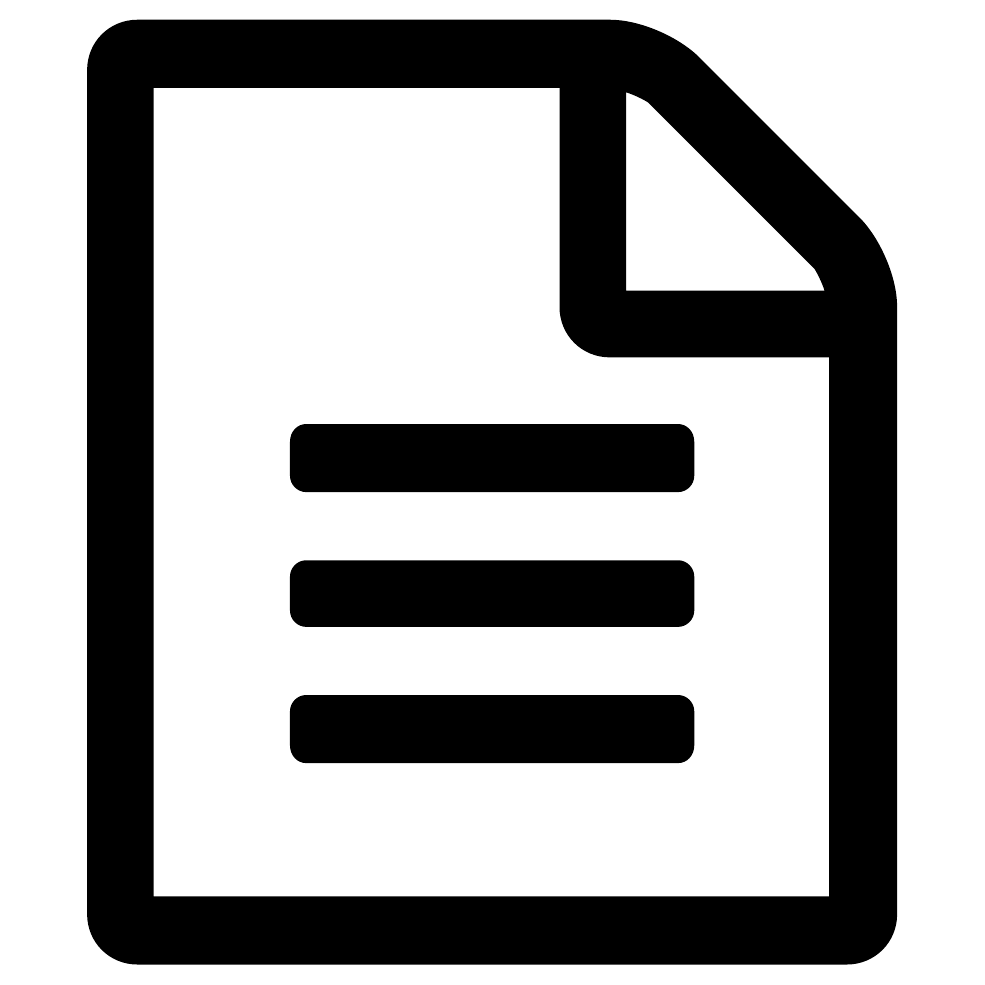}\xspace}
\newcommand\iconPointer{\includegraphics[width=0.01\textwidth,trim=0cm 2.0cm 0cm 0cm]{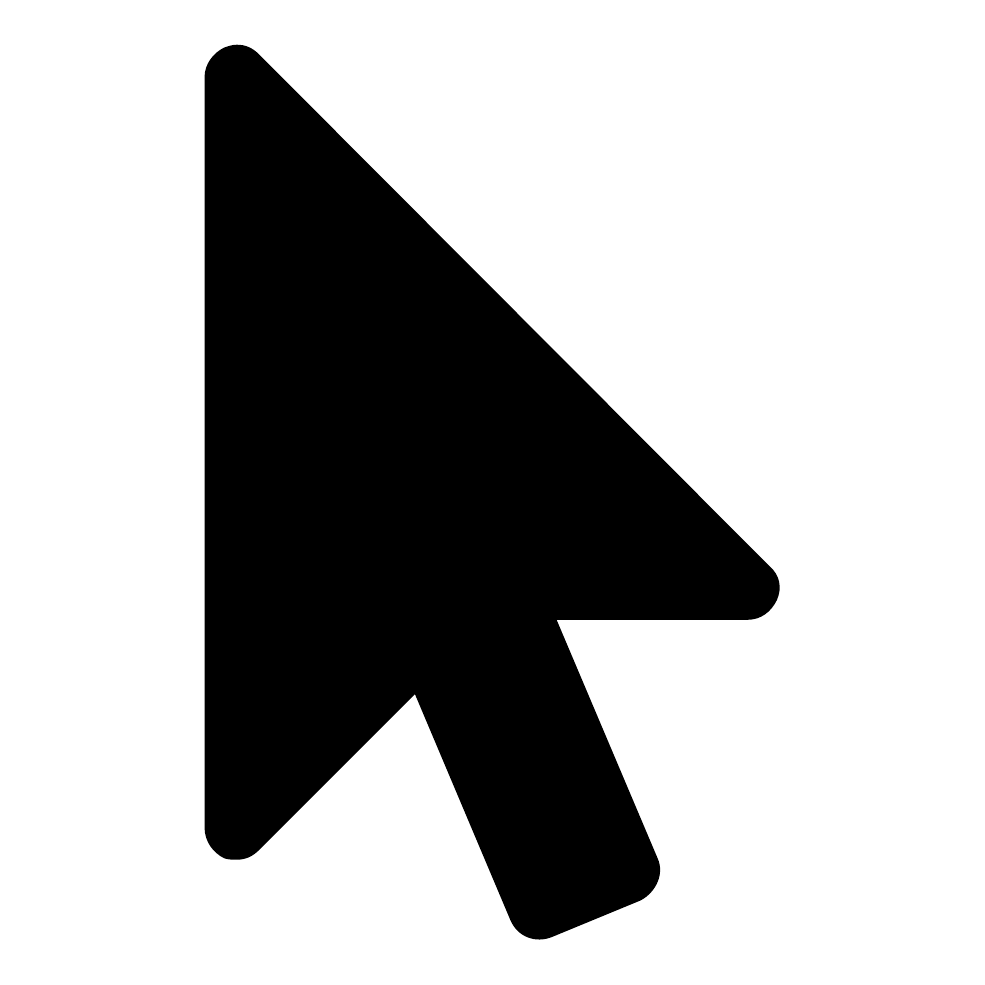}\xspace}
\newcommand\iconHook{\includegraphics[width=0.01\textwidth,trim=0cm 2.0cm 0cm 0cm]{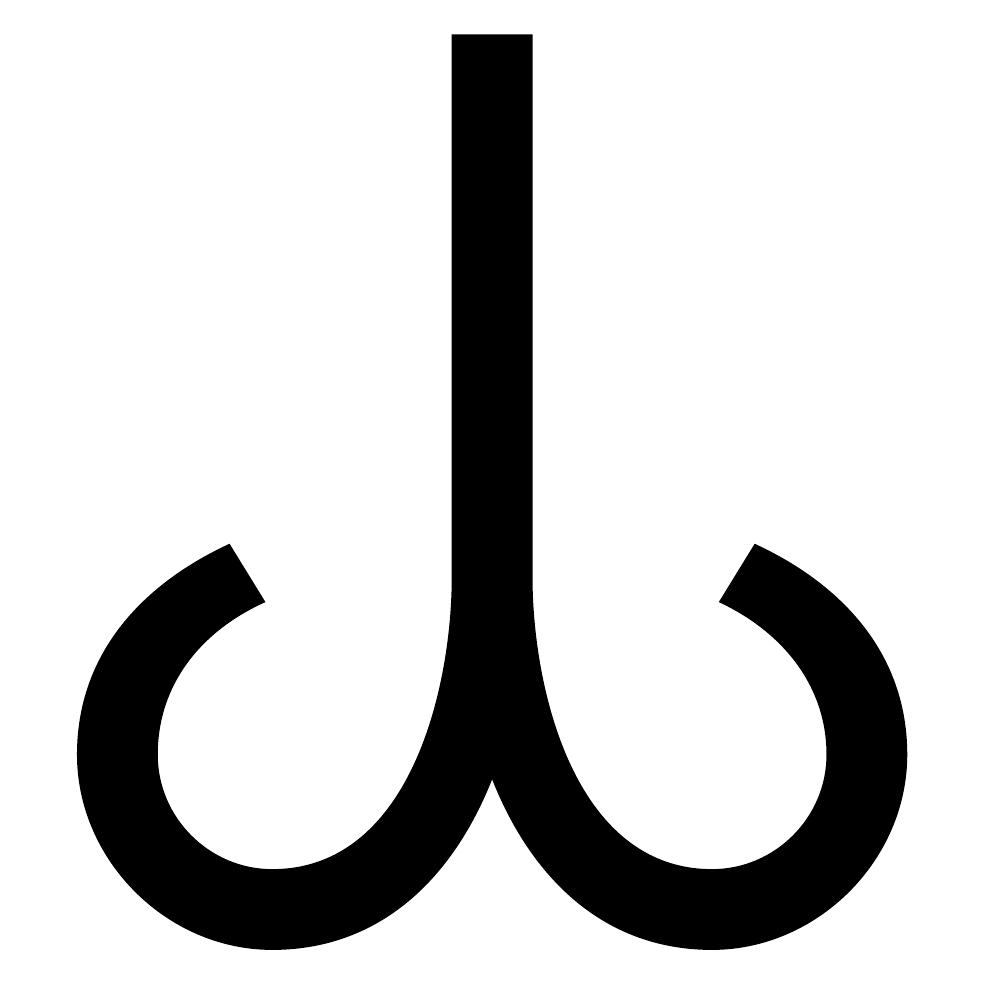}\xspace}

\newcommand\iconEyeText{\includegraphics[width=0.015\textwidth,trim=0cm 0.0cm 0cm 0cm]{icons/faEye}\xspace}
\newcommand\iconLightningText{\includegraphics[width=0.015\textwidth,trim=0cm 0.0cm 0cm 0cm]{icons/Lightning}\xspace}
\newcommand\iconStarText{\includegraphics[width=0.015\textwidth,trim=0cm 0.0cm 0cm 0cm]{icons/medstar}\xspace}
\newcommand\iconStopText{\includegraphics[width=0.015\textwidth,trim=0cm 0.0cm 0cm 0cm]{icons/stopsign}\xspace}
\newcommand\iconDisconnectText{\includegraphics[width=0.015\textwidth,trim=0cm 0.0cm 0cm 0cm]{icons/ztransf}\xspace}
\newcommand\iconTemplateText{\includegraphics[width=0.015\textwidth,trim=0cm 0.0cm 0cm 0cm]{icons/faFileText0}\xspace}
\newcommand\iconPointerText{\includegraphics[width=0.015\textwidth,trim=0cm 0.0cm 0cm 0cm]{icons/faMousePointer}\xspace}
\newcommand\iconHookText{\includegraphics[width=0.015\textwidth,trim=0cm 0.0cm 0cm 0cm]{icons/strictif}\xspace}

\newcommand{\CircleCustom}[4]{\raisebox{#3pt}{\CircledParamOpts{inner xsep=#1pt, inner ysep=#2pt}{1}{#4}}}
\newcommand{\CircleNumber}[1]{\Circled{#1}}

\newcommand\task[1]{#1:\xspace}

\title{Phish What You Wish}

\author{
\IEEEauthorblockN{Pascal Gadient, Pascal Gerig, Oscar Nierstrasz}
\IEEEauthorblockA{Software Composition Group, University of Bern\\Bern, Switzerland\\\href{http://scg.unibe.ch/staff/}{\faGlobe \hspace{0.1cm}scg.unibe.ch/staff}}
\and
\IEEEauthorblockN{Mohammad Ghafari}
\IEEEauthorblockA{School of Computer Science, University of Auckland\\Auckland, New Zealand\\\href{mailto:m.ghafari@auckland.ac.nz}{\faEnvelope \hspace{0.1cm}m.ghafari@auckland.ac.nz}}
}

\maketitle

\begin{abstract}
IT professionals have no simple tool to create phishing websites and raise the awareness of users.
We developed a prototype that can dynamically mimic websites by using enriched screenshots, which requires no additional programming experience and is simple to set up.
The generated websites are functional and remain up-to-date.
We found that 98\% of the hyperlinks in mimicked websites are functional with our tool, compared to 43\% with the best competitor, and only two participants suspected phishing attempts at the time they were performing tasks with our prototype.
This work intends to raise awareness for phishing attempts especially with local websites by providing an easy to use prototype to set up such phishing sites.
\end{abstract}


%===============================================================================
%=============================== NEW CHAPTER ===================================
%===============================================================================
\section{Introduction}
\label{sec:introduction}
Phishing is a social-engineering technique to collect sensitive information from people without their knowledge and consent.
The people who perform phishing, \ie the phishers are increasingly interested in lucrative high-profile targets aiming to steal intellectual property, corporate secrets, and sensitive information concerning national security~\cite{Hong:2012}.
They often deceive a victim to land on a fake website (\ie phishing website) through which they attempt to collect sensitive data.
There are numerous countermeasures to circumvent exposure to phishing, \eg browser extensions or DNS black-lists.
Nonetheless, it is one of the most common cybercrimes across the globe. For example, 65\% of businesses in the U.S. suffered from at least one successful phishing attack in 2019~\cite{ProofPoint:2020}.

In order to raise user awareness for these threats, several tools for instructors exist that can automatically create a replica of an original website.
For example, phishing kits are usually available online for well-known websites from shady sources, \ie a set of files including a template that mimics the design of the website being faked, server-side code to capture and send submitted data to the instructor, and optionally code to filter out unwanted traffic~\cite{Oest:2018}.
Moreover, rule-based phishing frameworks exist that patch the original HTML and JavaScript code on the fly according to predefined rules.
However, tools that modify original HTTP content must obey complex rules and thus work around existing security features found in web browsers, \eg they have to adjust or remove the provided Content-Security-Policy (CSP) information.
Therefore, they are very sensitive to changes in the original site and usually require updates when the original site or its infrastructure undergo some changes.
As a result, unique manual effort is still required for the majority of dynamic websites, because otherwise the phishing websites remain static, get out of date, or lack server-side features, which increases the likelihood that victims will identify their exposure to phishing.

In this work, we introduce a new technique inspired by \emph{browser isolation} technology\footnote{\url{https://www.secjuice.com/remote-browser-isolation-vendors/}} to automatically develop phishing websites that are, to a great extent, identical to their genuine counterparts.
For this technique, the phishing server acts as a proxy between a client browser and a server, but does not need any complex configuration.
The client component that is executed in a victim's web browser forwards every user interaction to the server, waits for the response, and then displays the received data of what we call an ``enriched image.''
An enriched image is a screenshot with overlaid interactive User Interface (UI) elements to provide a genuine web experience.
Considering this technique, we investigate the following two research questions:

\textbf{RQ$_{1}$}: \emph{Can we automatically create functional replicas of dynamic websites?}
We implemented a proof-of-concept prototype and evaluated it on the top ten most used websites according to \emph{Amazon Alexa}.
We found that 98\% of the hyperlinks in the mimicked websites are functional, compared to 43\% of the best competitor.
Moreover, to our knowledge, it is the first phishing framework supporting out of the box features that need interaction with the original server, ranging from a ``simple query recommendation while typing into a search box'' to a ``two-factor authentication (2FA) measure.''
For instance, we could log in to a major Swiss e-banking website that uses 2FA.

\textbf{RQ$_{2}$}: \emph{Do such replicas provide an authentic web experience?}
We asked 14 participants to perform five predefined tasks disguised as website usability studies.
The obtained results are encouraging.
From 42 page visits that involved phished websites generated by our prototype, only six (\ie 29\%) were considered suspicious, and a mere two participants suspected phishing attempts at the time they were using them.
Interestingly, most participants did not expect phishing attempts in the spoofed local site, instead they rather assumed an attack at the Facebook website, which was genuine.

In conclusion, we found that our prototype is much easier to work with and provides features unavailable in existing solutions:
\begin{inparaenum}[i)]
\item it is portable and requires no setup time,
\item users can optionally specify the domain name they would like to phish without requiring any additional parameters or templates, 
\item thanks to its proxy architecture, it captures every user interaction and can trace a victim throughout an entire browsing session, and store all sensitive inputs,
\item the content of a phished website is always in sync with the original website,
\item it supports various features to create a genuine browsing experience, \eg the browsing history, adapted URL paths, and favicons, 
\item our implementation can circumvent CSP protective measures, and finally, 
\item our implementation offers support for common captchas and 2FA mechanisms.
\end{inparaenum}
As a result, we see much potential for our approach to be used for anti-phishing trainings, and obviously, an increased need for using a variety of anti-phishing techniques in order to mitigate such threats.

The remainder of this paper is structured as follows.
We discuss phishing attacks in~\autoref{sec:related-work} and present the concept behind our prototype in~\autoref{sec:phishing-on-demand}, where we also compare existing phishing frameworks.
We present our evaluation regarding the functionality and the authenticity of the website replicas in~\autoref{sec:experiences-in-practice}.
Next, we elaborate on the limitations and mitigations in~\autoref{sec:limitations-mitigations-opportunities}.
Finally, we report the threats to validity in~\autoref{sec:threats-to-validity} and conclude in~\autoref{sec:conclusion}.

%===============================================================================
%=============================== NEW CHAPTER ===================================
%===============================================================================
\section{Phishing Attacks}
\label{sec:related-work}
In~\autoref{SUBSEC:offensive-measures} we present works, which provide a state-of-the-art overview of existing techniques used by phishers, and in~\autoref{SUBSEC:general-measures} we discuss corresponding anti-phishing measures.

\subsection{Modus Operandi}
\label{SUBSEC:offensive-measures}

Simple static phishing kits for the masses are still very prevalent and originate from only a few sources.\footnote{\url{https://www.imperva.com/blog/our-analysis-of-1019-phishing-kits/}}
Such phishing kits are deployed to servers under direct control of phishers, to free hosting services~\cite{Moore:2007}, to hijacked servers that host one or more legitimate website~\cite{Han:2016}, to public clouds~\cite{Han:2015}, and even to botnets~\cite{McGrath:2008}.
As a covert measure, 95\% of the deployed phishing kits are enforcing \code{htaccess} rules that block ``unwanted'' visitors as Oest~\etal discovered when they examined 1\,794 live phishing kits from 2016 through mid-2017~\cite{Oest:2018}.
Hosting providers are particularly at risk, because more than 95\% of them do not even run an anti-virus scan with up-to-date signatures once a month~\cite{Canali:2013}.
Such a lax behavior encourages criminals to run privileged escalation attacks that might yield access to other websites published on the same server that could be modified and misused for phishing as well.
When a phishing kit that targets a specific website is deployed, the manually applied modifications are rather minor as Cui~\etal found when they compared the DOMs (Document Object Models) of different phishing sites~\cite{Cui:2018}, \eg adversaries add an input form, or replace a few images or some text.

Rule-based phishing frameworks provide more genuine experiences, but they still require time demanding manual work before they can operate.
Such tools usually modify website traffic with the help of regular expressions that match relevant code.
Typical modifications are the replacement of hard-coded URLs and the removal of security-related HTTP headers.
\emph{Modlishka}, for example, is a powerful open-source HTTP proxy that is able to apply website code changes on the fly.\footnote{\url{https://github.com/drk1wi/Modlishka}}
The required code changes for a certain URL are specified in the corresponding website template.
Currently, there exists one for \emph{Microsoft Office 365} and one for \emph{Google GSuite}.
Furthermore, there exists a very popular proxy implementation called \emph{evilginx2}, which includes templates for fourteen other major websites.\footnote{\url{https://github.com/kgretzky/evilginx2}}

Adopting a ready-to-deploy phishing kit also raises threats to its users.
Cova~\etal observed that criminals started to introduce in their kits hidden backdoors that transmit the collected data not only to its users, but also to third parties, \eg its originator~\cite{Cova:2008}.
McCalley~\etal show that such hidden backdoors have evolved using different email address encoding techniques, \eg hexadecimal string representations or custom array-related code~\cite{McCalley:2011}.
Moore~\etal investigated the re-compromising of internet hosts and found that by performing Google searches using versioning information of phishing kits, attackers discover hijacked and still vulnerable phishing hosts ready to take additional phishing kits~\cite{Moore:2009}.
Finally, Birk~\etal reveal that organized cyber-crime grants custom requests for phishing kits~\cite{Birk:2007}, however it is not clear if such individual solutions contain backdoors.

\subsection{Anti-phishing Measures}
\label{SUBSEC:general-measures}
A large body of research has focused on connection-based anti-phishing strategies.
Hong~\etal use several lexical features of existing phishing URLs, curated in PhishTank,\footnote{\url{https://www.phishtank.com/}} 
such as an URL's length or the number of special symbols to build a classifier that detects potential phishing URLs~\cite{Hong:2020}.
They benchmark different machine learning classification algorithms on their results, \eg variants of SVM and random forest.
There exist similar works that use other features such as the age of the domain in combination with the WHOIS record~\cite{Jain:2018} or the presence of HTTPS~\cite{Rao:2020}.
There also exist similar works that use other classification algorithms, \eg deep neural networks~\cite{Kp:2020}, or a combination of them~\cite{Zamir:2020}.
Joshi~\etal describe a browser plug-in that uses the received HTTPS certificate to hash the entered password before it is submitted~\cite{Joshi:2009}.
When used with forged URLs, the adversary consequently only captures garbage data due to a certificate mismatch.
Fortunately, many of these techniques are available in practice, \eg by using the ``Google Safe Browsing API''\footnote{\url{https://safebrowsing.google.com}} or the ``Kaspersky Protection Plugin.''

Researchers have also proposed techniques that detect phishing websites from the content (text, code, rendered images) of websites. For instance,
Zhang~\etal implemented a text-based classifier that extracts, among other data, a website's five most frequent words and sends them to Google to verify if the returned results match the domain in question~\cite{Zhang:2007}.
Liu~\etal use the HTML source code to extract the layout of a website to which they apply an image segmentation algorithm~\cite{Liu:2006}.
They compare, among other data, the resulting image segments between benign and malignant websites and received high similarity scores for corresponding pages, \eg for eBay's official website and its phished counterpart.
Chiew~\etal extract company logo images from a website under test and process them with Google's image search to conclude whether a page is phished~\cite{Chiew:2015}.
They assume that an image and the corresponding website are legitimate when its URL matches one of the top rated URLs returned by Google.
Hara~\etal compare the screenshot of a suspicious website with genuine screenshots that they already have in a database~\cite{Hara:2009}.

Finally, industry has tried to overcome phishing with external authentication hardware and software.
The \emph{W3C WebAuthn} web standard specifies an interface for web applications that offers public-key cryptography.\footnote{\url{https://www.w3.org/TR/webauthn/}}
Compliant websites can request authorization that lets the browser ask for a second authentication factor, \eg a USB or NFC token.
There exist other similar concepts like \emph{Mobile-ID}, which leverages SMS messages to establish a secure context~\cite{Bicakci:2014}.
External authenticators can also be used to measure timing differences as Ulqinaku~\etal describe with \emph{2FA-PP}, a system that evaluates the network's round-trip time to the server to detect phishing attempts~\cite{Ulqinaku:2019}.
If implemented properly, such mechanisms effectively protect users from phishing threats.

%===============================================================================
%=============================== NEW CHAPTER ===================================
%===============================================================================
\section{Screenshot-based Phishing}
\label{sec:phishing-on-demand}
We present a prototype to create phishing websites that behave, to a great extent, identically to their genuine counterparts.
The prototype can collect sensitive data or trace an entire browsing session, and working with it is very easy:
a phishing website can be deployed within seconds.
In the following we explain the concept, we discuss key features, and we share technical details of a prototype that we have created.
The source code and a one-click executable for all major platforms is publicly available on \emph{GitHub}.\footnote{\url{https://github.com/pgadient/PhishWhatYouWish}}

\begin{figure*}
\centering
\includegraphics[scale=0.66,trim=1.5cm 7.2cm 1cm 7.2cm]{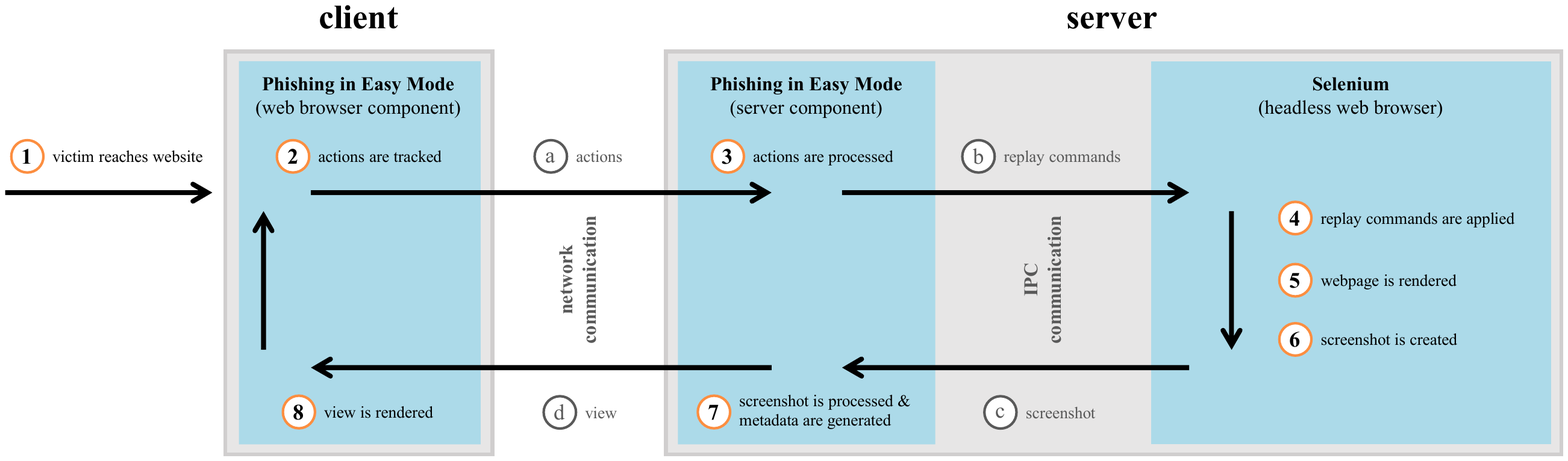}
\caption{Responsibilities of the client and the server}
\label{fig:data-flow}
\end{figure*}

\subsection{Process}
The responsibilities of the client and the server are revealed in~\autoref{fig:data-flow}.
The grey rectangles show the involved devices, \ie the victim's client computer on the left and the adversary's server on the right.
The blue rectangles illustrate the executed applications, \ie the web browser started by the victim, and the Java server daemon as well as the headless web browser started by the adversary.
The arrows indicate the different steps.
The steps that describe an action are labeled by numbers, whereas transmitted data are labeled by letters.

We discuss each step in detail.
\CircleNumber{1}~A victim must be tricked into opening a website operated by our prototype.
This step is identical to existing techniques where attackers convince victims to open a certain URL, for example, by email, instant messaging, or phone.
If the victim's browser requests such a URL, the server component quickly spawns a headless browser instance with the pre-configured website, takes a screenshot and sends it back to the client attached to the web browser component, which now becomes active on the client.
\CircleNumber{2}~The delivered JavaScript code tracks the victim's mouse and keyboard actions and sends them through the internet to the Java-based server daemon.
\CircleCustom{6.4}{6.4}{1.25}{a}~The transmitted actions include mouse click coordinates, keyboard hits, changes in text boxes, and page navigation events.
\CircleNumber{3}~The server daemon decodes the received actions and generates replay commands for the headless web browser following the same order.
The commands are sent by using \emph{Inter-Process Communication (IPC)}.
\CircleCustom{4}{4}{0.25}{b}~The forwarded replay commands might look like ``execute click at point (52/142)'', or ``press the keys \code{J}, \code{o}, \code{h}, \code{n}.''
\CircleNumber{4}~The headless web browser maintained by Selenium,\footnote{a comprehensive browser automation toolkit, \url{https://selenium.dev/}} executes the collected replay actions in the same order they are received.
\CircleNumber{5}~The web browser starts to re-render the website based on the provided actions.
The rendering itself can be performed off-screen, \ie hidden in memory without the use of any visible UI elements to enable support for arbitrarily large screen resolutions at clients.
\CircleNumber{6}~When the configured time-out expires, the headless browser takes a screenshot of the rendered website and sends it back to the server daemon by IPC.
\CircleCustom{6.4}{6.4}{1.25}{c}~The screenshot is encoded in an image format understood by modern browsers.
\CircleNumber{7}~The server daemon processes the screenshot, and traverses the headless browser's DOM of the previously rendered website to find and extract embedded UI elements.
The resulting view is then transmitted over the internet back to the client.
\CircleCustom{4}{4}{0}{d}~The view consists of the screenshot, the location and content of text boxes or buttons as well as hyperlinks.
\CircleNumber{8}~The JavaScript code executed on the client receives the view, decodes the data, and renders it within the client's browser ensuring a seamless update between each cycle.

\subsection{Advantages}
Our implementation provides several features unseen in existing implementations.
We discuss each feature and explain how traditional template-based methods and our prototype support them.
We rely for information about static template-based methods on the two comprehensive open-source projects \emph{Gophish}\footnote{\url{https://getgophish.com}} and \emph{Phishing Frenzy},\footnote{\url{https://www.phishingfrenzy.com}} for information about rule-based methods on the two open-source projects \emph{evilginx2} and \emph{Modlishka}, and for information about handcrafted methods on our own expertise.
Please note that the opportunities offered by handcrafted websites can be partially transferred to template-based methods due to their support for manual code interventions.
We expect that traditional phishing pages are created by using these techniques, based on the findings of Cova~\etal who performed a comprehensive study of more than 500 phishing frameworks found in the wild~\cite{Cova:2008}.
They mention that phishing kits usually comprise two types of files: the files to display a copy of the targeted website, and the scripts used to save the phished information, and send it to criminals.
For each aspect, we argue that a handcrafted solution would require more time due to its inherent complexity, however more flexibility is generally achieved.
An experienced developer might be necessary depending on the desired quality of the implementation.

\subsubsection{Accuracy}
Accuracy refers to the visual similarity of a phished page compared to the original.
If the replica is accurate, no differences can be observed by a potential victim.
\Template barely support websites that use dynamic content, \eg overlays created with JavaScript, because templates cannot automatically adapt to such variable content.
\Phod, instead, presents screenshots of web pages which are by definition an accurate representation of what users of the websites would see in their browsers.

\subsubsection{Back-end Logic}
Back-end logic is an umbrella term for code executed on a server that is protected from public access, \ie it cannot be downloaded or repurposed.
Typical use cases that require back-end logic are search results, user profiles, and authorization, \eg with 2FA.
\Template cannot replicate back-end logic without complex manual intervention.
\Phod, on the contrary, does not require any back-end logic treatment or circumvention, because it interacts like every web browser does.

\subsubsection{Browser Plug-in Support}
Browser plug-ins serve numerous purposes, \eg they can provide ad blocking, export videos, keep notes, remember credentials, \etc
Plug-ins can interact with pages, and therefore manipulate a site's content transparently to the potential victim.
This provides new opportunities to circumvent image detection approaches, \eg by blocking ads, or by injecting on the fly some fake banner overlays.
\Template do support fundamental changes on original pages, however they must be implemented manually instead of using existing browser plug-ins.
\Phod has full control over the rendered HTML and JavaScript code, and it supports Mozilla Firefox plug-ins out of the box.
For that reason, our system facilitates arbitrary changes on web pages by directly injecting code into the website, by applying browser plug-ins, \eg ad blockers and dark themes, or by manipulating the screenshot pixel-wise.

\subsubsection{Completeness}
Completeness refers to the provided functionality compared to the original page.
If a website is incomplete, victims might become suspicious.
Relative resource paths are particularly problematic for completeness, since their path only remains valid on the original server, but not on the phishing server.
Therefore, either the paths must be adjusted to the resources on the original server, if possible, or all resources must be tracked and copied to the phishing server.
\Template try to find and fix problematic resource paths automatically, but they might fail when assembled at run time or obfuscated in code.
A developer can define exceptions for problematic code.
\Phod does not require any changes because the website's origin is not altered in any way.

\subsubsection{Collected Data}
Phishing purposefully requires sensitive data to be collected, \eg user names, passwords, tokens, \etc
The more data that can be exfiltrated, the better the chances are for an adversary to successfully exploit people. 
\Template induce constraints on the possible collection of data, \ie one is supposed to use predefined methods for gathering data.
\Phod does not require any operations for gathering data as the exfiltration process remains transparent to the victim.
All key presses, mouse click coordinates, and optionally all created website screenshots are captured.

\subsubsection{Continuous Tracking}
Continuous tracking enables observation of users throughout their entire surf sessions.
The continuous data stream that web surfers generate is very valuable.
For instance, someone could record personal browsing preferences and use them for blackmailing, or collect multiple logins to gather access to different services.
Nevertheless, many existing phishing sites hand over control to the real site after the phished data has been gathered.
\Template allow multiple phishing pages to be spawned on the same domain, but they cannot serve arbitrary pages on demand.
They would require prior setup of every page a user could visit, which is unfeasible.
\Phod does not require any website preparation, hence it supports the logging of entire surf sessions beyond domain boundaries.
For example, if a search engine is set as the landing page, the user can be tracked while entering a search term and choosing an entry from the results, and even while navigating through the desired web pages.

\subsubsection{Recency}
Recency refers to actuality of phished sites, \ie whether a phished site reflects arbitrary changes of the original, and how long it takes to reflect them.
The more recent a replica is, the more convincing it generally can be for a potential victim.
\Template are incapable of automatically integrating arbitrary changes into phished sites that are already live.
For that reason, the process has to be repeated every time the original website is modified.
\Phod always provides the latest version of a website to the potential victim.

\subsubsection{Required Knowledge}
Some IT knowledge is required in order to run a phishing campaign.
The less knowledge is required, the more people can perform phishing.
\Template, such as phishing kits, only require basic web development knowledge, \eg the ability to install software and to import an existing website into a web framework.
The import process can be either performed manually by copy-pasting code into the framework or by importing relevant pages through an assistant that will guide the user through required changes.
There exists some pre-baked code that performs the extraction of credentials,\footnote{\url{https://github.com/mgeeky/PhishingPost}} however various manual adjustments can still be required.
Proxy-based solutions assume a deep understanding of web development to create the mandatory configuration templates.
\Phod requires only basic IT knowledge, but no web development skills:
the only required configuration data is the URL to phish.

\subsubsection{Required Set-up Time}
The required set-up time indicates the time it takes to set up a phishing instance.
The efficiency increases by providing a shorter setup, \ie more time remains available for other tasks.
\Template require the installation of one or more frameworks, but in general, they try to reduce user interventions wherever possible.
They achieve that goal by providing predefined capture schemes that can be applied to original pages, or by providing predefined templates from which a user can choose.
Nevertheless, a user must read the documentation to understand all the features required to successfully set up a phishing page.
\Phod requires only very little time, \ie a few seconds to adjust the URLs of the website that should become replicated and to start the application.

We believe that screenshots are superior to existing video stream based solutions,\footnote{\url{https://www.browserling.com/}}\footnote{\url{https://github.com/i5ik/ViewFinder}} because very often 
\begin{inparaenum}[i)]
\item the interaction response time of locally rendered screenshots is much lower than of videos from remote machines, \eg the scrolling experience is much smoother,
\item the clarity is much higher, because the views can use little to no compression, which is not suitable for video streams.
Therefore, users cannot see any additional compression artifacts even when they quickly scroll through content.
\end{inparaenum}

Furthermore, we expect that our prototype can be useful in practice for different stakeholders:
\emph{End users} might use this tool to safely browse the internet from a remote machine.
\emph{Software developers} might use this tool to easily capture screenshots from a website for their documentation.
\emph{IT security staff} might use this tool to educate people about the threats of phishing and to assess them when a more genuine phishing experience is required compared to a virtual machine running a remote desktop session host.
Our prototype does not require manual work to set up phishing sites for these use cases, unlike existing tools.

\subsection{Implementation}
We implemented a prototype that leverages a client/server architecture:
a single portable Java-based application contains all the required code to run the server.
We did not rely on existing web rendering proxy frameworks, because they could not offer all the basic features we required, \eg access to the website's DOM.

If started by double-clicking the executable file, the application will spawn two web servers and one headless browser component.
One web server is responsible for the delivery of static resources to the client, \ie the HTML and JavaScript code that establishes the initial connection to the server.
The other web server opens two WebSocket connections that begin to listen for clients; one WebSocket connection is used for the transmission of screenshots, the other for the bidirectional transmission of commands, \eg instructing the client to forward to a given page, or notifying the server about a key press event from the user.
After these servers are online, the application spawns a \emph{Selenium} instance that itself starts a remote-controlled headless Firefox browser.
The application is ready when the instrumented browser is awaiting any inputs.

The website on the client tracks the victim's actions, submits them to the server, and renders screenshots with the help of metadata received from the server.
In more detail, the victim's keystrokes, mouse click coordinates, and changes in text boxes are tracked and submitted to the server with custom JavaScript code.
Similarly, right mouse click events can be transmitted to the server, however additional client side code would be required to render a genuine context menu depending on the operating system and regional settings.
The use of the HTML \texttt{canvas} element instead of the \code{image} tag for displaying screenshots ensures smooth image transitions for the client without any artifacts or other visual glitches, \eg flickering.
The metadata received from the server is processed as follows:
the client separates coordinates and embedded text of text boxes, the locations of buttons, and clickable areas of hyperlinks.
Next, text boxes and buttons that contain the original text, as well as clickable hyperlinks, are rendered locally on top of the screenshot.
The reason for this sophisticated implementation is the authentic look and feel that locally rendered GUI elements provide:
they enable user interactions without any delays, provide a consistent user experience across other non-phished websites, and they enable support for mobile devices with on-screen keyboards.
Even more, built-in password managers detect the text boxes and can propose credentials.

The server replays the victim's keystrokes and mouse clicks on the original page by using a scriptable headless web browser that provides an API for instrumentation.
Due to the superior documentation and better feature support, we chose the Mozilla Firefox web browser over Google Chrome.
After each replayed user action, a new screenshot of the resulting page is sent back to the client.

Several additional features have been implemented on both ends to facilitate a more realistic surfing experience for the victim:
\begin{inparaenum}[i)]
\item the title of the original page is mimicked by using the data within the HTML \texttt{<title>} tag,
\item the original website icon, also known as ``favicon,'' is copied from the original location to the server where it is provided to the client, 
\item the URL's path is replicated according to the original structure, \eg \texttt{https://accounts.google.com/login} results in \texttt{http://phi\-sh\-ingsite.com/login},
\item copy-paste events into text boxes with keyboard short-cuts are captured,
\item browser history with favicons is supported and enables the forward and backward navigation,
\item Google Captchas are supported although they occasionally take more time to complete than usual,
\item drag support for sliders is provided to bypass most slider protections, and finally, 
\item optional ad blocking is supported, since ad-loaded pages can take considerably longer to load, and thus would negatively impact the victim's experience.
\end{inparaenum}

%===============================================================================
%=============================== NEW CHAPTER ===================================
%===============================================================================
\section{Evaluation of the Website Replicas}
\label{sec:experiences-in-practice}
For the evaluation of the website replicas generated by our tool we were particularly interested in their functionality, \ie are the replicas usable, and in their authenticity, \ie are the replicas convincing to people.
We try to answer the two research questions in the remainder of this section.

\subsection{Functionality}
In this subsection we aim to answer \textbf{RQ$_{1}$}: \emph{Can we automatically create functional replicas of dynamic websites?}
There exist numerous rule-based open-source phishing tools that claim they can successfully phish popular websites.\footnote{\url{https://github.com/search?q=phishing+proxy&type=Repositories}}
\emph{Modlishka}, for example, even claims that it works on most websites without any templates.
To validate such claims, we tested two typical and very popular representatives, \ie \emph{evilginx2}\footnote{commit: fe4e3431430df7d9c283493fd1a58196026acfb2} and \emph{Modlishka},\footnote{commit: 9d57cfb5c01a84554eb06497b304444ff28226d7} against the top ten most visited websites reported by Amazon Alexa, a well-known website traffic statistic aggregator.
All of these websites contain dynamic elements and change frequently, \eg to announce discounts.
Because most of those sites do not maintain the login page on the same URL, we also included their corresponding login sites in our URL test set since every top ten site has at least one.
Finally, we apply our prototype to the same URLs and compare the results.

We closely followed the referenced guides on project pages to set up the server-side applications.
We used for \emph{evilginx2} the Facebook template and for \emph{Modlishka} always the Google template to improve its performance even when working with different URLs, because those two templates were the only ones that matched an entry in our top ten list.
Specifying no template at all would decrease the quality of phishing, because only basic replacement rules could be applied to web traffic, \eg replacing the originating URL with the phishing domain.
We used default settings except for our prototype where we disabled the ad block plug-in to obtain more comparable results.
For \emph{evilginx2} we created an HTTPS certificate from the \emph{Mozilla Let's Encrypt} initiative, and for \emph{Modlishka} we used the generated self-signed certificate which we manually added to our browser's trusted certificate store.
No certificate was required by our prototype, because it currently only supports plain text HTTP traffic between the victim and the phishing server.
On the client-side, we used \emph{Mozilla Firefox} 76.0.1 64-bits, a display with a resolution of 1920 by 1200 pixels, and we maximized the browser windows.

For each website listed in the Amazon Alexa top 10 ranking, we opened the phished versions of the website and investigated each clickable element, \ie a button, hyperlink, or scripted content sensitive to a mouse click.
For each clickable element, we assessed the element itself (link visibility and function) as well as its target (visual glitches).
We only considered clickable elements at the top level which are directly visible without any further user interaction.
We noted for each clickable element whether it works correctly or it suffers from an issue, \ie we always assigned one of five states:
\iconStarText denotes a link that works as expected and leads to a site without noticeable UI glitches.
\iconEyeText denotes a link with visual glitches or that leads to a site that suffers from visual glitches but in any case still works as expected.
\iconLightningText denotes a link that does not work as expected, \ie is invisible or without function, or leads to a site that is broken, \eg target site is not reachable or blank.
\iconStopText denotes a link to a site in which an HTTP Content-Security-Policy (CSP) warning is triggered.
\iconDisconnectText denotes a link that leads to a drop-out, \ie a site off-limits to the phishing proxy, usually a genuine HTTPS site.
We further introduced \iconPointerText to denote the number of clickable elements on the site, \iconTemplateText to denote the used phishing site template, and finally, \iconHookText with the values ``yes'' or ``no'' to report whether the phishing attempt was successful.

\begin{table*}
\centering
\caption{Phishing results for the global top ten websites}
\label{tab:top-10-results}
\begin{tabular}{ m{0.6cm} m{3.2cm} m{0.8cm} m{2.2cm} m{1.8cm} m{1.8cm} m{1.8cm} }
\multicolumn{1}{c}{\textbf{Rank}} & \multicolumn{1}{l}{\textbf{Website}} & \multicolumn{1}{c}{\Large\iconPointer\normalsize} & \multicolumn{1}{c}{\textbf{Legend}} & \multicolumn{1}{c}{\textbf{evilginx2}} & \multicolumn{1}{c}{\textbf{Modlishka}} & \multicolumn{1}{c}{\textbf{\Phod}} \\ \hline
\thead{\\1\\} & \thead[cl]{\\google.com\\} & \thead{\\17\\} & \thead{\iconTemplate\\\iconStar~ / \iconEye~ / \iconLightning\\\iconStop~ / \iconDisconnect} & \thead{n/a\\-/-/-\\-/-} & \thead{Google\\15/0/0\\1/1} & \thead{-\\14/3/0\\0/0} \\
\thead{\\2\\} & \thead[cl]{\\youtube.com\\} & \thead{\\37\\} & \thead{\iconTemplate\\\iconStar~ / \iconEye~ / \iconLightning\\\iconStop~ / \iconDisconnect} & \thead{n/a\\-/-/-\\-/-} & \thead{Google\\1/28/8\\0/0} & \thead{-\\36/1/0\\0/0} \\
\thead{\\n/a\\} & \thead[cl]{\\accounts.google.com\\} & \thead{\\8\\} & \thead{\iconTemplate\\\iconStar~ / \iconEye~ / \iconLightning\\\iconStop~ / \iconDisconnect~ / \iconHook} & \thead{n/a\\-/-/-\\-/-/\textbf{no}} & \thead{Google\\4/0/3\\1/0/\textbf{no}} & \thead{-\\8/0/0\\0/0/\textbf{no}} \\ \hline
\thead{\\3\\} & \thead[cl]{\\tmall.com\\} & \thead{\\64\\} & \thead{\iconTemplate\\\iconStar~ / \iconEye~ / \iconLightning\\\iconStop~ / \iconDisconnect} & \thead{n/a\\-/-/-\\-/-} & \thead{Google\\0/0/0\\0/64} & \thead{-\\40/23/1\\0/0} \\
\thead{\\8\\} & \thead[cl]{\\login.tmall.com\\} & \thead{\\43\\} & \thead{\iconTemplate\\\iconStar~ / \iconEye~ / \iconLightning\\\iconStop~ / \iconDisconnect~ / \iconHook} & \thead{n/a\\-/-/-\\-/-/\textbf{no}} & \thead{Google\\4/0/0\\0/39/\textbf{no}} & \thead{-\\16/23/4\\0/0/\textbf{no}} \\ \hline
\thead{\\4\\} & \thead[cl]{\\facebook.com\\} & \thead{\thead{54\\(47)}} & \thead{\iconTemplate\\\iconStar~ / \iconEye~ / \iconLightning\\\iconStop~ / \iconDisconnect~ / \iconHook} & \thead{Facebook\\18/13/1\\0/15/\textbf{yes}} & \thead{Google\\39/1/2\\0/12/\textbf{no}} & \thead{-\\39/15/0\\0/0/\textbf{yes}} \\ \hline
\thead{\\5\\} & \thead[cl]{\\qq.com\\} & \thead{\\110\\} & \thead{\iconTemplate\\\iconStar~ / \iconEye~ / \iconLightning\\\iconStop~ / \iconDisconnect} & \thead{n/a\\-/-/-\\-/-} & \thead{Google\\73/0/37\\0/0} & \thead{-\\61/49/0\\0/0} \\
\thead{\\n/a\\} & \thead[cl]{\\mail.qq.com\\} & \thead{\\13\\} & \thead{\iconTemplate\\\iconStar~ / \iconEye~ / \iconLightning\\\iconStop~ / \iconDisconnect~ / \iconHook} & \thead{n/a\\-/-/-\\-/-/\textbf{no}} & \thead{Google\\8/0/5\\0/0/\textbf{no}} & \thead{-\\10/2/1\\0/0/\textbf{yes}} \\ \hline
\thead{\\6\\} & \thead[cl]{\\baidu.com\\} & \thead{\\30\\} & \thead{\iconTemplate\\\iconStar~ / \iconEye~ / \iconLightning\\\iconStop~ / \iconDisconnect~ / \iconHook} & \thead{n/a\\-/-/-\\-/-/\textbf{no}} & \thead{Google\\9/0/9\\0/12/\textbf{no}} & \thead{-\\24/6/0\\0/0/\textbf{yes}} \\ \hline
\thead{\\7\\} & \thead[cl]{\\sohu.com\\} & \thead{\\95\\} & \thead{\iconTemplate\\\iconStar~ / \iconEye~ / \iconLightning\\\iconStop~ / \iconDisconnect~ / \iconHook} & \thead{n/a\\-/-/-\\-/-/\textbf{no}} & \thead{Google\\39/7/20\\0/29/\textbf{no}} & \thead{-\\59/36/0\\0/0/\textbf{yes}} \\
\thead{\\n/a\\} & \thead[cl]{\\mail.sohu.com\\} & \thead{\\8\\} & \thead{\iconTemplate\\\iconStar~ / \iconEye~ / \iconLightning\\\iconStop~ / \iconDisconnect~ / \iconHook} & \thead{n/a\\-/-/-\\-/-/\textbf{no}} & \thead{Google\\4/0/1\\0/3/\textbf{no}} & \thead{-\\8/0/0\\0/0/\textbf{yes}} \\ \hline
\thead{\\9\\} & \thead[cl]{\\taobao.com\\} & \thead{\\53\\} & \thead{\iconTemplate\\\iconStar~ / \iconEye~ / \iconLightning\\\iconStop~ / \iconDisconnect} & \thead{n/a\\-/-/-\\-/-} & \thead{Google\\2/0/5\\0/46} & \thead{-\\47/6/0\\0/0} \\
\thead{\\n/a\\} & \thead[cl]{\\login.taobao.com\\} & \thead{\\52\\} & \thead{\iconTemplate\\\iconStar~ / \iconEye~ / \iconLightning\\\iconStop~ / \iconDisconnect~ / \iconHook} & \thead{n/a\\-/-/-\\-/-/\textbf{no}} & \thead{Google\\14/0/37\\0/1/\textbf{no}} & \thead{-\\24/27/1\\0/0/\textbf{yes}} \\ \hline
\thead{\\10\\} & \thead[cl]{\\360.cn\\} & \thead{\\36\\} & \thead{\iconTemplate\\\iconStar~ / \iconEye~ / \iconLightning\\\iconStop~ / \iconDisconnect} & \thead{n/a\\-/-/-\\-/-} & \thead{Google\\18/0/12\\0/6} & \thead{-\\17/14/5\\0/0} \\
\thead{\\n/a\\} & \thead[cl]{\\i.360.cn/login\\} & \thead{\\15\\} & \thead{\iconTemplate\\\iconStar~ / \iconEye~ / \iconLightning\\\iconStop~ / \iconDisconnect~ / \iconHook} & \thead{n/a\\-/-/-\\-/-/\textbf{no}} & \thead{Google\\4/2/4\\0/5/\textbf{no}} & \thead{-\\13/1/1\\0/0/\textbf{yes}} \\ \hline \hline
 & \thead[cl]{\thead{\thead{\textbf{Summary (absolute)}}}} & \thead{\thead{\thead{635}}} & \thead{\iconStar~ / \iconEye~ / \iconLightning\\\iconStop~ / \iconDisconnect~ / \iconHook} & \thead{\thead{18/13/1\\0/15/1}} & \thead{\thead{234/38/143\\2/218/0}} & \thead{\thead{416/206/13\\0/0/7}} \\
 & \thead[cl]{\thead{\thead{\textbf{Summary (relative)}}}} & \thead{\thead{\thead{100\%}}} & \thead{\iconStar~ / \iconEye~ / \iconLightning\\\iconStop~ / \iconDisconnect~ / \iconHook} & \thead{3\%/2\%/0\%\\0\%/2\%/11\%} & \thead{37\%/6\%/23\%\\0\%/34\%/0\%} & \thead{66\%/32\%/2\%\\0\%/0\%/78\%} \\\\
\end{tabular}
\end{table*}

In the remainder of this subsection we discuss the results with respect to our observations in~\autoref{tab:top-10-results}.

\subsubsection{Templates}
We can see that phishing templates are unavailable for most of the top ten sites.
In other words, templates for Chinese websites are completely missing.
Therefore, such websites break compatibility with existing tools and are mostly unsupported.
For example, \emph{evilginx2} is unable to spoof anything without a template.
Our prototype does not require any template.

\subsubsection{Clickables}
Using a template, \emph{evilginx2} can successfully spoof websites, but some clickables might not work as it seems that dynamic functionality is reduced from the original website to ease development of the template.
Nevertheless, most intra-domain links worked as intended (56\%) or suffered only from minor visual issues (41\%).
\emph{Modlishka}, on the other hand, can work without specific templates and successfully spoof many pages.
Surprisingly, the Google template did not fully work:
\emph{Modlishka} could successfully spoof the login screen, but the ``Next'' button did not act properly which made the phish impossible.
We conjecture that the current website underwent changes which are incompatible with the template.
In general, intra-domain links either worked successfully (56\%) or not at all (34\%), \ie the browser reported for many clickables the ``no connection could be established'' error.
Only a minority suffered from visual glitches (9\%).
In contrast, the spoofing attempts by our tool are mostly successful (66\%) or suffer from minor glitches (32\%).
Only a minority failed (2\%) mostly due to missing support for captchas that require drag-and-drop mouse support (not yet implemented at the time of the evaluation) or detection mechanisms for instrumented browser instances, \ie Google stated ``we detected an unsupported browser.''

\subsubsection{Content-Security-Policy (CSP)}
CSP is an HTTP header feature to restrict origins for web resource requests.
If a website implements CSP properly, requests to protected resources from foreign (phishing) domains are blocked.
We encountered such CSP warning messages in websites from Google when we used \emph{Modlishka}.
Our prototype cannot suffer from such issues, because the code is always rendered in its original domain context.

\subsubsection{Tracking}
Every victim that follows a genuine HTTPS link outside the adversary's domain is lost and cannot be controlled anymore.
Consequently, we can say that the fewer times a drop-out occurs, the more likely it is to gather sensitive data from a victim.
In general, many links to external sites remain unchanged, \ie 32\% (\emph{evilginx2} on Facebook) or 34\% (\emph{Modlishka}) of all clickables lead to a loss of control over the victim.
We can observe that more drop-outs occur when no matching template is available.
With existing tools, drop-outs can even be desired to avoid detection when template support for the requested site is incomplete.
In contrast, our prototype tracks a user effortlessly throughout different sites.
As a result, no drop-outs can occur.

\subsubsection{Data Collection}
After a successful spoof the proxy must collect the relevant sensitive data to achieve a successful phish.
For existing proxies, templates are a major contributing factor for a successful phish, because templates specify which pieces of a website contain the sensitive information and need special treatment.
\emph{evilginx2} can successfully capture entered data, but only for \emph{Facebook} which results in an 11\% success rate for our list of login sites.
According to our experiments, \emph{Modlishka} does not know what to capture from the HTML traffic when no matching template is available.
Hence, it cannot capture desired data in any of the tested pages, \ie it achieves a success rate of 0\%.
This is completely different with our prototype.
Since it captures all user input and views, a seamless reconstruction or takeover of entire browsing sessions becomes feasible.
Therefore, it achieves a success rate of 78\%, \ie seven times more than the second best tool we tested.

As we found, contrary to the claims of existing tools, it is possible to automatically spoof some sites, but phishing still remains a complex task, which requires manual effort.
In particular, existing tools have problems with complex scripts, frequent website code changes, captchas, or security and QR codes from external services that break their functionality.
Although our prototype lacks some basic functionality, \eg listeners for function keys, it works far better than any other tested approach.
If we would add the clickables with visual glitches, which are still usable, to the successfully phished clickables (66\% + 32\%), a success rate of almost 98\% could be achieved which is more than twice as high compared to the results from the second best tool, \ie \emph{Modlishka}  with 43\% (37\% + 6\%).
Based on our experiences from the user study, we expect that a broken design usually does not alarm people since such glitches can also occur on regular websites when certain elements do not load properly, however broken functionality raises suspicion, because such behavior is rather uncommon when no error message is visible.
Moreover, our prototype is resilient against CSP protections and drop-outs, and it does not require any templating.

%===============================================================================
%=============================== NEW CHAPTER ===================================
%===============================================================================
\subsection{Authenticity}
\label{sec:empirical_study}
In this subsection we aim to answer \textbf{RQ$_{2}$}: \emph{Do such replicas provide an authentic web experience?}
For that purpose, we performed an empirical phishing study that we disguised as a website usability study to not raise any suspicions amongst the participants.

\begin{figure}
\centering
\includegraphics[scale=0.183,trim=0cm 3cm 0cm 0cm]{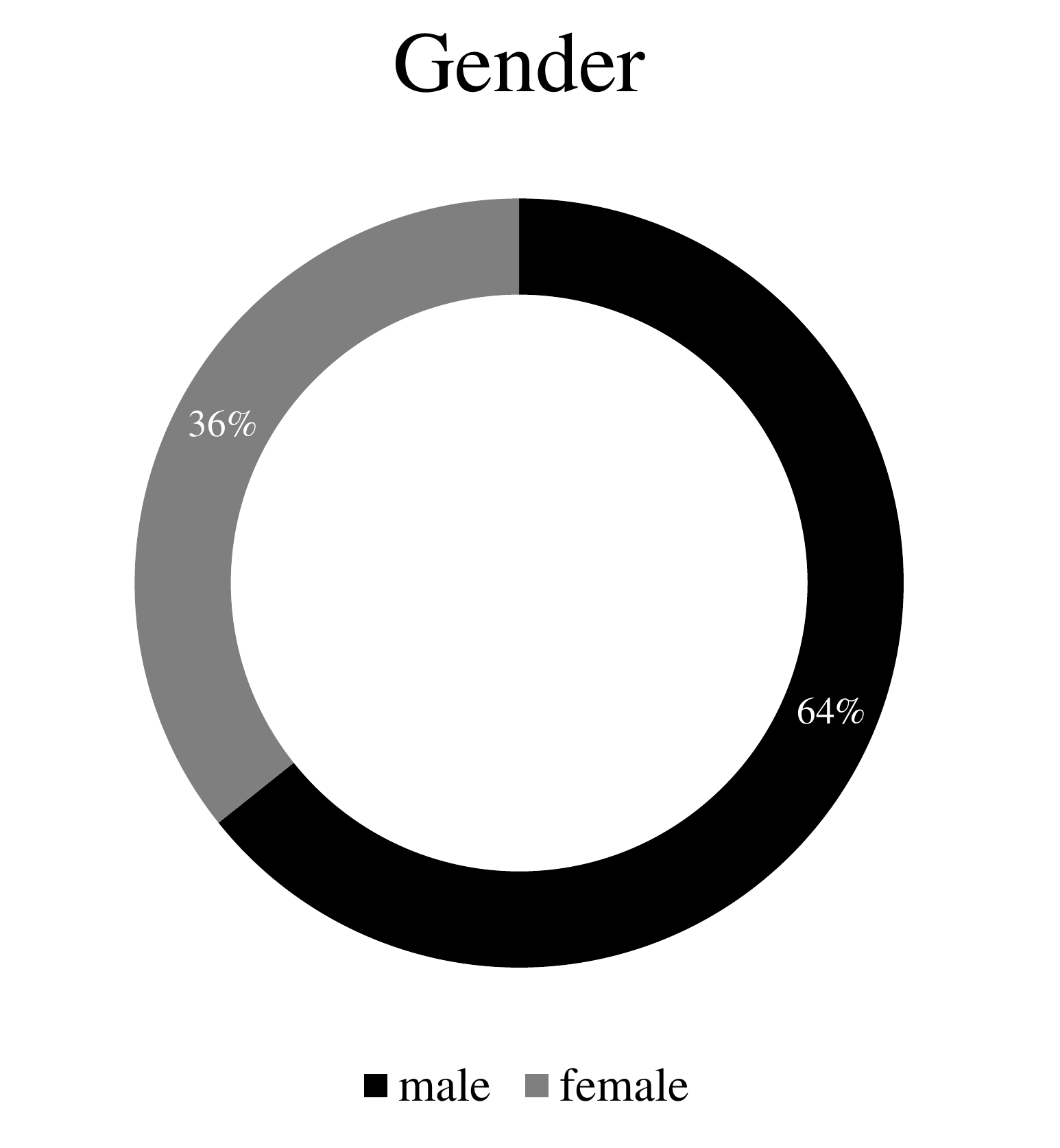}
\includegraphics[scale=0.183,trim=0cm 3cm 0cm 0cm]{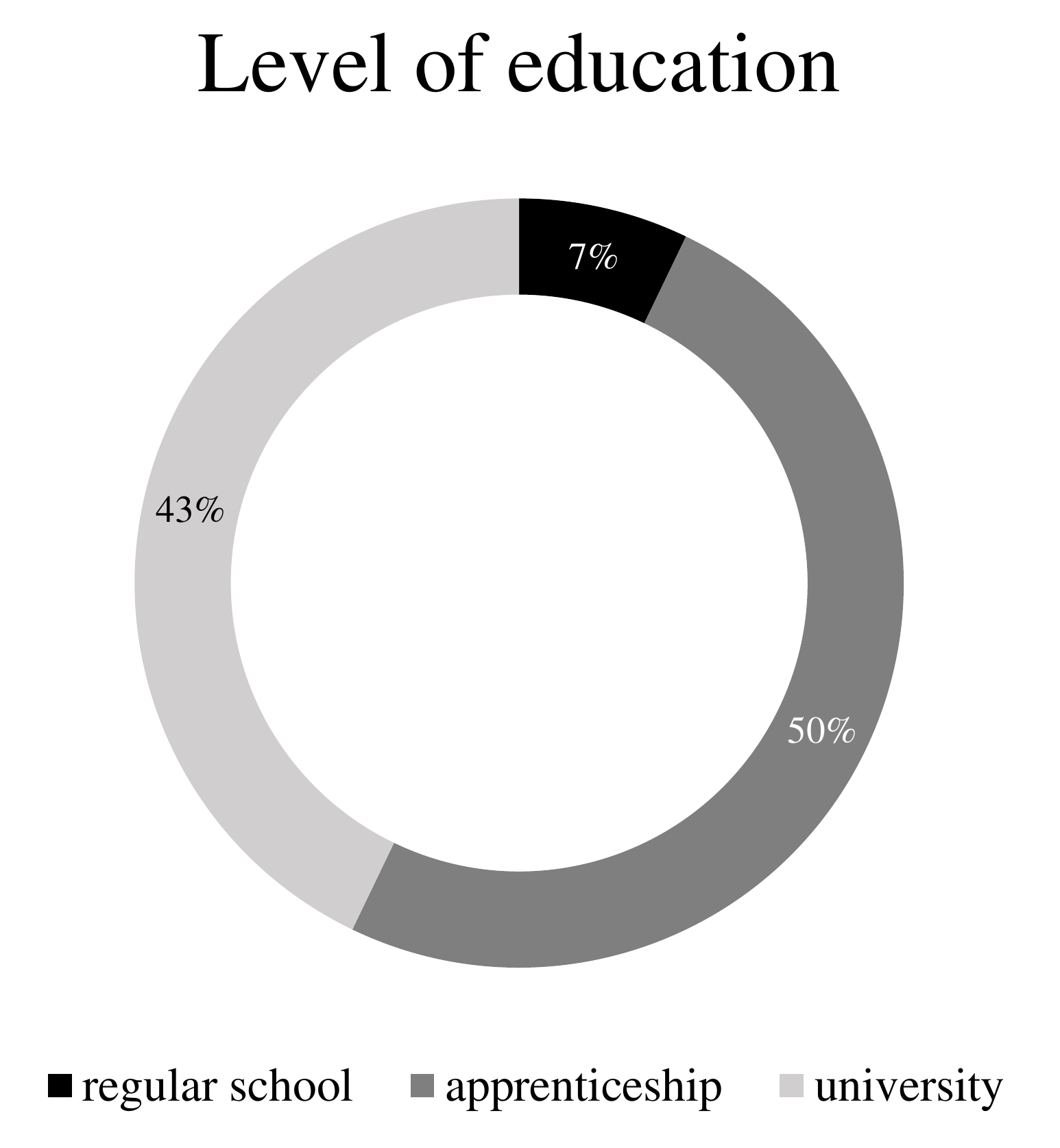}
\includegraphics[scale=0.183,trim=0cm 3cm 0cm 0cm]{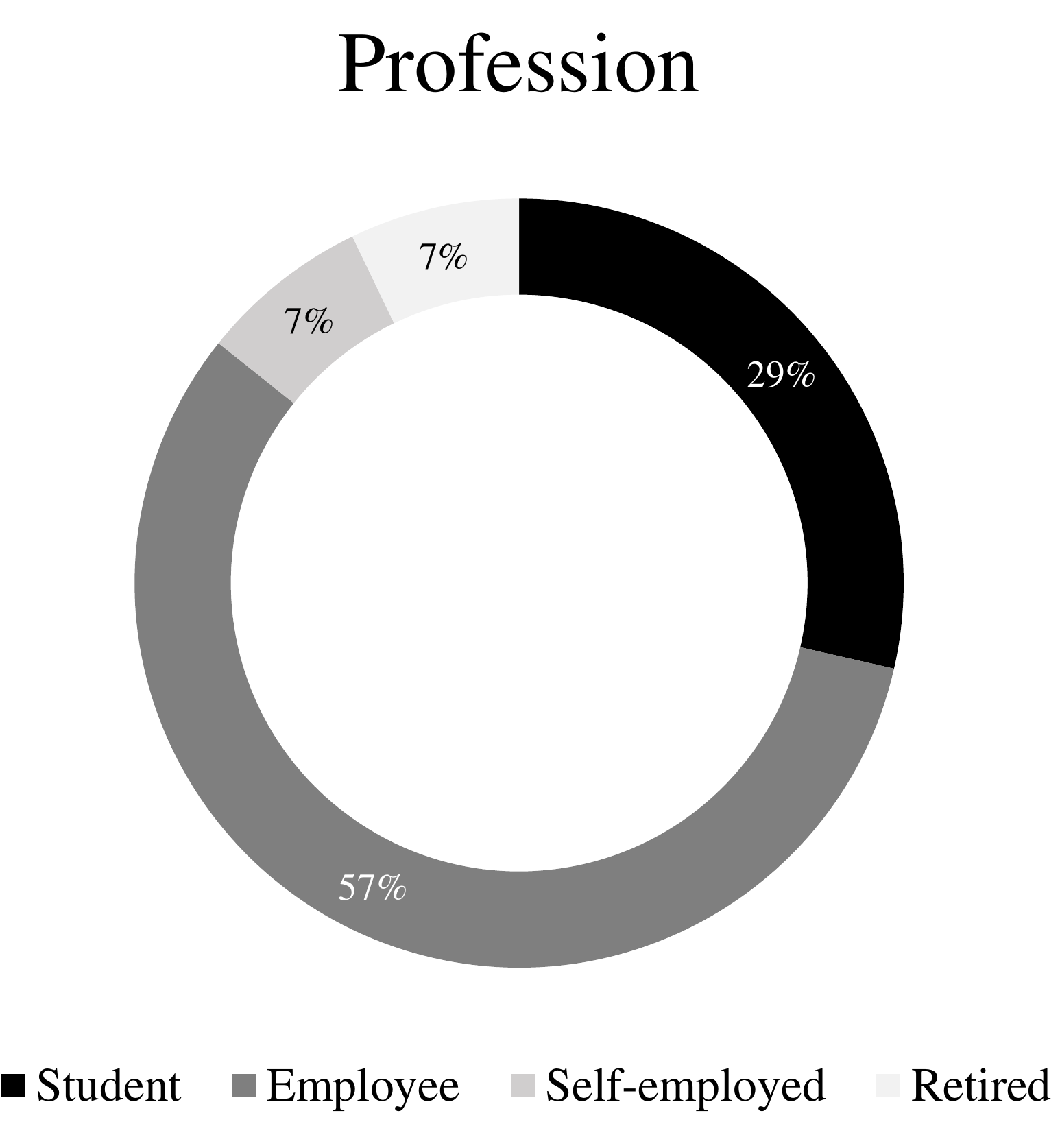}
\caption{Demography of the participants}
\label{fig:demography}
\end{figure}

\subsubsection{Demography}
The study involves 14 participants with diverse backgrounds and from different age groups, \ie from computer science students to retired business owners.
The participants were conveniently acquired from the university, family, friends, and workplace.
As shown in~\autoref{fig:demography}, five participants were female and nine male, the level of education was for one participant regular school, for seven apprenticeship, and for six a university, and the current profession was for four being a student, while eight were employed by a company, one was self-employed, and finally, one was already retired.
On average, the participants had 14 years of web experience, and two of them had some web development experience, \ie one nine years and another six months.
Two participants already suffered from a phishing attack, one when he was selling collectible cards online, and another one installed a virus through a phished page.
Luckily, none of them suffered from any financial losses caused by phishing so far.

\subsubsection{Questionnaire}
The questionnaire we used comprises three sections and can be found in the public \emph{Github} repository, which is referenced in~\autoref{sec:phishing-on-demand}.
In the first section, we gathered personal information, \eg level of education, profession, or experience using the web.
In the second section, we let the participants interact with different websites and after each interaction session we questioned them about their experience, the usability, and other observations they drew.
In the last section, we gathered more information regarding the familiarity with the previously visited sites, and at last, to not draw any attention during the study, their knowledge of phishing.
If they lacked some experience with phishing, we explained what it is, and finally, we questioned them in retrospect for each task the perceived likelihood the relevant site was phished.

\subsubsection{Performing the tasks}
Every participant had to solve five tasks, where each task focuses on one particular kind of website, \ie originals without any modifications, originals replicated by our prototype, or traditional phishing websites.\footnote{\url{https://github.com/ashanahw/Gmail_Phishing}}
Hence, the tasks differ in the use of tools, \ie one task did not involve any tool, one involved a traditional phishing tool, and three involved our prototype.
In terms of complexity, one task was only about visual inspection, one about navigation through websites, and three about entering credentials into login pages.
Except for the first introductory visual inspection task, the order of the tasks has been randomized to mitigate any potential bias in the results.
We only enforced a time limit for the first Google task and the task with the local website to avoid unnecessarily long survey sessions, because there was no specific goal to achieve for the participants.
After a participant reached the maximum allowed time, we took control of their browser and continued with the evaluation.
In order to let the participants mainly focus on the website itself, all tasks were performed in full screen mode so that the address bar is invisible.
While the supervisor was preparing a task, the participants were not allowed to watch the computer screen.
In this work we closely follow ethical principles.
In all our experiments we used credentials that we solely created for the sake of phishing.
No participant had to provide personal data.

The individual tasks were as follows.
\begin{inparaenum}[1)]
\item \task{Visual inspection of a static website with our prototype}
each participant should look at \href{https://www.google.com/}{Google Search} for a maximum of ten seconds.
Our intention was to see whether the participants notice a difference in the generated screenshot compared to the original website.
\item \task{Static website with our prototype}
on Google Search, each participant had to search for \href{https://www.srf.ch/}{``SRF (Swiss Radio and Television),''} to navigate to that page, and they were encouraged to continue exploring that website for up to 30 seconds.
Our intention was to see whether our tool can cope with complex dynamic websites, and how participants perceive phishing on local sites.
\item \task{Login website with our prototype}
each participant was tasked to search for \href{https://www.amazon.com/}{``Amazon''}, to navigate to that page, and finally, to log in with the given credentials.
Our intention was to see whether our tool can cope with complex dynamic websites, and how participants perceive phishing on international sites.
\item \task{Login site without any modifications}
each participant was told to search for \href{https://www.facebook.com/}{``Facebook''}, to navigate to that page, and finally, to log in with the given credentials.
Our intention was to see how participants perceive phishing on international genuine sites that are well-known as phishing targets.
\item \task{Login site with a traditional phishing tool}
each participant had to log in through the presented Google GMail login website by using the provided credentials.
Our intention was to see whether the participants notice the phishing attempt with an existing off-the-shelf phishing kit that targets an international company.
This is the only task that involves a functional phishing kit, because the acquisition of such kits is not trivial.
Unfortunately, we could not replicate this task with our tool since Google disallows access to their log-in site for headless browsers.
\end{inparaenum}

Immediately after completion of each task, we asked every participant the following five questions:
\begin{inparaenum}[i)]
\item Do you have any experience with the visited website, and if so, what kind of experience?
\item What was the website's level of usability on a scale from one to five, where one refers to very low and five refers to very high?
\item Was there something you really liked concerning the usability?
\item Was there something you really disliked concerning the usability?
\item Is there something that could be considered to improve usability?
\end{inparaenum}

\subsubsection{Findings}
We discuss the feedback from the participants, where we specifically focus on their perception of phished websites.
The received feedback matches the three categories, usability, authenticity, and further observations.

\subsubsubsection{Usability}
We asked all participants after every task to rate the perceived usability level on a Likert-scale between one (very low) and five (very high).
The participants rated the usability on average with a score of 3.9 (\ie high) for the original web page and 3.6 (\ie medium to high) for the web page from a traditional phishing tool.
When we average the usability scores of the other three tasks that used our prototype, we see a score of 3.6 (\ie medium to high) and the most prevalent score was 4 (high), which has been reported in 17 cases.

In the traditional phishing website, some embedded hyperlinks and the focus order are broken.
Moreover, after stealing the credentials, the website tries to redirect the user to the original website, which requires the participants to log in a second time.
One participant noticed the focus order bug, and another one noticed that this website was quicker than our prototype.
Surprisingly, only four participants were irritated that the credentials had to be entered twice, and two of them were not even expecting a phishing attempt.
However, these problems only slightly reduced the perceived usability for most participants.

Based on the participants' responses for our prototype, the lower usability score has two main reasons:
First, sometimes our tool produced unexpected results, and second, our tool introduced a page load delay depending on the complexity of the requested website.
Although our tool works with most of the existing websites, it still suffers from occasional glitches, \eg suddenly appearing or misplaced text boxes, and problems with the infinite scrolling feature used by websites that quickly increases the size of the screenshots and, consequently, also the website response time.
The additional delay before a website is displayed has also been reported:
three participants criticized the longer than usual loading time once they clicked on a hyperlink to another website, whereas one participant preferred that the results of his \emph{Google Search} were displayed all at once, \ie no changes to the page layout occurred during the rendering process.
This is a result of the simplified rendering on the client that only needs to show one static image instead of thousands of characters and HTML elements manipulated by JavaScript code at run time.

We conclude that the loading time of a web page has a significant impact on how people rate a page's usability, but differences in usability levels do not make people suspect any malicious activity, except when people have to enter their credentials more than once.

\subsubsubsection{Authenticity}
After the completing experiment the participants were asked to judge the likelihood of the seen websites being phished replicas, again by using a Likert-scale from one (very unlikely to be a phishing site) to five (very likely to be a phishing site).
We received for sites from our prototype an average score of 2, whereas the original Facebook page received 2.3 and the traditional phishing page achieved 2.4.
Interestingly, the average scores for the Amazon and Facebook website were up to 0.8 points higher compared to the local site and Google.
We expect that to be a result of the reputable protection offered by Google, and the lack of phishing awareness for regional sites.
The most prevalent score for replicas from our prototype was 1 (very unlikely), which has been reported in 16 cases (41\%).
If we consider that a failed phishing attempt has a score of 3 (likely a phishing page) or higher, from 42 phishing attempts that involved websites generated by our prototype, only 12 (\ie 29\%) attempts were failing.

The majority of our screenshots were the most authentic to participants, and our replicas were barely distinguishable from the originals.
To our surprise, the original Facebook was rated more likely to be phished than pages that were indeed phished.
We can only explain this by the many phishing scams that became public in the news, thus drawing attention to this specific website.
Furthermore, the traditional \emph{Google Gmail} phishing site used in our experiment did not receive the highest Likert-score, instead the Amazon replica of our tool did.

Many of the participants were rather experienced in using Google and Facebook, \ie 11 participants use Google multiple times a day, whereas 3 participants use Facebook daily and 6 at least once a week.
The design of the traditional \emph{Google Gmail} phishing site we used in our experiment dates back to March 2017, therefore the style clearly differs from recent iterations.
No participant, not even the one who uses this service on a daily basis noticed that, even when entering the credentials for the second time on a page with a different design.
On the contrary, one participant considered a task spoofed, because URLs on Google Search's result page were not green like on his own computer.
However, this assumption is incorrect, because Google Search recently updated their design.

In conclusion, the participants did not pay much attention to the design of websites they use.
Moreover, based on our results they seem to associate the term ``phishing'' with particular websites of large companies that are well-known for their involuntary involvement in phishing scams, rather than with sites of small- to medium-sized businesses.
Hence, they more likely report phishing activity on those websites, even when no attacks are performed.
Most participants genuinely did not know how to decide if a site is phished or not.

\subsubsubsection{Further Observations}
One participant felt responsible for not completing a task even though the application did not correctly render the screenshots.
Another participant mentioned that the traditional phishing site started to look phishy when ``unexpected messages'' appeared in the browser, \eg a password save dialogue box should not appear on regularly visited sites.
Finally, one participant did not expect any phishing on a site shown by our prototype, because there was no login dialogue on the viewed page.

%===============================================================================
%=============================== NEW CHAPTER ===================================
%===============================================================================
\section{Limitations and Mitigations}
\label{sec:limitations-mitigations-opportunities}
In this section we discuss the prototype's limitations and potential mitigation strategies.

\subsection{Limitations}
The tool currently suffers from inherent limitations such as delays or limited interactivity on dynamic content for websites that heavily use scripting.
Nevertheless, these problems can be mitigated with additional engineering effort.

\subsubsection{Computational Delays}
For our implementation we use a simple architecture that introduces only little processing overhead while still enabling the rapid development of features.
Nevertheless, further improvements would include hardware acceleration support, \eg for image compression, and more use of low-level programming languages, \eg C or C++.
Next, on the proxy server, every website request must be at least partially finished before a screenshot can be sent back to the client, whereas a regular client can start displaying a website right away as soon as data is received.
For this problem, a solution could be to use video streams instead of screenshots similar to the implementation used in game streaming services.
By using such technology, the user would see every step of the website rendering which is even more similar to the regular browsing experience.
Moreover, the server needs to fully execute every incoming website request.
Because complex websites require much processing and RAM for display, \eg up to several hundred megabytes for a single page, the server hardware must be rather powerful.
Load-balancing features that would reduce the load on a single server instance have not yet been implemented.

\subsubsection{Domain Use}
The evaluation, reasonable selection, and registration of domain names is essential for successful phishing campaigns.
However, our prototype does not support domain-related tasks.

\subsubsection{Transmission Delays}
Our prototype uses a man-in-the-middle concept which reflects an image-based HTTP proxy for the internet.
Inherent to this architecture, the required network traffic increases in most cases substantially, \ie screenshots of websites usually are in the range of megabytes of data compared to text and code that remain rather in the kilobytes.
Generally, this increase in data that needs to be transmitted to the client can increase the time required to display a website.
This problem can be reduced by using high-bandwidth network interconnects for the server.
Another approach would be to use a better image compression algorithm than \emph{Portable Network Graphics (PNG)} that could massively reduce the amount of transmitted data.
However, this change might also reduce the visual quality of the screenshots since PNG uses a lossless algorithm.

\subsubsection{Limited Interactivity}
The screenshot update frequency is limited by the CPU capabilities of the server and the network interconnect.
As a result, dynamic elements such as CSS animations and videos will be transformed into static images and therefore might begin to stutter depending on the image size.
However, on average a single client connected to an off-the-shelf computer that runs our prototype can achieve between 20 and 30 screenshots per second when the website scales to the full HD resolution, \eg \url{www.google.com}.
Currently, we have no solution for that particular problem, but dynamically transmitting only fractions of the image, or accessing the framebuffer of the graphics card bypassing Selenium APIs might reduce the problem.

\subsubsection{Phishing Campaigns}
Professional phishing tools allow one to create and send phishing emails to individuals or address groups.
They provide mail templates and provide placeholders for personal formulations, \eg the victim's name, which are then automatically replaced before sending the message.
In that regard, functionality to import large data sets from Comma-Separated Value (CSV) files is present.
For instance, such CSV files can contain victims' email and postal addresses, phone numbers, and countries of residence.
Moreover, they collect statistics, \eg the time a phishing email has been dropped, when the email has been opened by the victim, or if and when the contained hyperlink has been accessed.
To check if an email has been opened, the tools usually embed a small invisible unique image, \ie tracking pixel, that is requested from the phishing server at the time the email is presented to the victim.
Our prototype is not intended to support phishing campaigns, \ie we neither support phishing emails, nor do we leverage any statistics about a campaign's success rate.

\subsection{Mitigation Strategies}
Anti-phishing measures aiming at the detection of text, source code, or website images have no impact on our prototype, because these file-based resources are not used in their original form.
As a result, we see four options to mitigate this phishing threat.

\subsubsection{Data Flow Analysis}
Our approach forwards all user input to the network socket.
This behavior could be leveraged to decide whether the website has a malicious intent.
Since the browser has full control over user input, JavaScript execution, and network communication, it could label data entered by a user, and track the labels across the DOM and even through the JavaScript execution engine.\footnote{a browser extension that provides taint analysis of string values in JavaScript, \url{https://github.com/ollseg/ttt-ext}}
If the browser detects that a website continuously sends labeled data over the network, it should display a warning message to the user.
Revoking the access to said site might be inappropriate, because some benign websites can reveal such behavior, \eg multiplayer games.

\subsubsection{Domain Analysis}
Our implementation provides exact visual replicas of websites, but their domain names might still differ.
This situation can be leveraged by a detection tool.
Suppose that a user already possesses credentials to be vulnerable to a phishing attempt.
Consequently, the user already had to visit certain benign websites in order to create a user account on them.
During that process, the browser could start to ``know'' what the original websites look like, \eg by building a ground-truth dataset based on screenshots.
If later a website screenshot looks similar to one from the ground-truth dataset but uses a different domain name, the browser should block access to the site and show a warning message.

\subsubsection{WebAuthn}
Authentication using external hardware is currently not supported by our prototype, because the server has no direct access to the client's hardware interfaces.

\subsubsection{Detection of Browser Instrumentation}
The headless browser instance uses plug-ins that enable remote control.
Such plug-ins can be detected by websites, for example, as performed by Google's authentication website, which prevents successful phishing with our tool.

%===============================================================================
%=============================== NEW CHAPTER ===================================
%===============================================================================
\section{Threats to Validity}
\label{sec:threats-to-validity}
A major threat to validity is the selection and number of the participants in this study, \ie whether our selection of participants is representative of the public.
We strived to include people with various backgrounds and internet knowledge.
Moreover, we gathered demographic information to see if we can find any correlations.

As with any Likert-scale, we cannot guarantee the accuracy and comparability of the participants' responses.
On principle, these responses are subjective and can be influenced by their feeling at that time.
In addition, the experiment might have induced stress in some people, which further impacts the outcome.

Since our implementation is not bug-free, some bugs did occur during a few experiments.
Although we immediately took countermeasures, this could still influence their response.

We did not perform the experiment with every web page in the internet, but only five.
However, we tried to select websites with high impact that are supported by most tools to receive reasonable results.
The success rate of a phishing campaign also depends on the quality of the used domain name.
However, the domain registration process is out of scope for this work.

We might have misinterpreted or misunderstood responses of the participants.
We tried to mitigate that problem by reviewing our notes after the experiment with every participant.
We did not yet implement and validate our proposed optimizations, hence we can only hypothesize about their benefits and compare the expected effects with other well-researched implementations found in popular software.

There is a threat to construct validity through potential bias in our expectancy.

%===============================================================================
%=============================== NEW CHAPTER ===================================
%===============================================================================
\section{Conclusion}
\label{sec:conclusion}
The creation of phishing websites often requires expensive manual work even with the help of tools.
Therefore, scammers are primarily attracted by major international websites with a large reach.
In consequence, IT professionals do not have access to an effective tool to raise the awareness of phishing in their companies, and moreover, people seem to underestimate the potential phishing threat from local sites.
We have explored a potential solution to these problems, \ie we developed a prototype that can dynamically mimic websites by using enriched screenshots, which requires no additional programming experience and is simple to set up.
We found that 98\% of the hyperlinks in mimicked websites are functional with our tool, compared to 43\% with the best competitor.
Moreover, only 29\% of the page visits from 14 participants were considered as suspicious, and only two participants suspected phishing attempts at the time they were performing their tasks.
We believe that our open-sourced tool has value for different stakeholders, and that this threat requires more attention, especially when considering the emerging ultra-broadband network technologies, \ie fiber landline and 5G cellular networks.

\section*{Acknowledgment}
We thank Sebastiano Panichella for his feedback on the survey design, and Yannick Hänni for the further improvement of the tool and his assistance in the user study.
We gratefully acknowledge the financial support of the Swiss National Science Foundation for the project
``Agile Software Assistance'' (SNSF project No.\ 200020-181973, Feb.\ 1, 2019 - April 30, 2022).

\balance

\bibliographystyle{IEEEtran}
\bibliography{PhishingOnDemand}

\end{document}

%% file: preamble.tex
\usepackage{textcomp}
\usepackage{amsthm}
\usepackage{xspace}
\usepackage{ifthen}
\usepackage{amsmath}
\usepackage{amssymb}
\usepackage{balance} % for balancing last page reference column
\usepackage{listings}
\usepackage[normalem]{ulem} %emphasize still italic
\usepackage{array, booktabs}
\usepackage[pdftex,colorlinks=true,pdfstartview=FitV,linkcolor=black,citecolor=black,urlcolor=black]{hyperref}
\usepackage[most]{tcolorbox}
\usepackage{paralist} %to list extends of previous work
\usepackage{fontawesome}      % for icon next to scg URL next to author list on title page
\usepackage{adjustbox} 		% for vertical text in tables
\usepackage{pifont} 		% for crosses and ticks in tables
\usepackage{circledsteps}	% for circled numbered steps in text, corresponding .sty file must reside in same folder
\usepackage{makecell} % for line breaks within cells in tables
\newcolumntype{R}[2]{
    >{\adjustbox{angle=#1,lap=\width-(#2),margin*=0.4em 0em 0em 0em}\bgroup}
    l
    <{\egroup}
}
\let\oldFootnote\footnote
\newcommand\nextToken\relax
\renewcommand\footnote[1]{%
    \oldFootnote{#1}\futurelet\nextToken\isFootnote}
\newcommand\isFootnote{%
    \ifx\footnote\nextToken\textsuperscript{,}\fi}

\newcommand{\id}[1]{$-$Id: scgPaper.tex 32478 2010-04-29 09:11:32Z oscar $-$}

\newcommand{\ie}{\emph{i.e.},\xspace}
\newcommand{\eg}{\emph{e.g.},\xspace}
\newcommand{\etal}{\emph{et al.}\xspace}
\newcommand{\etc}{\emph{etc.}\xspace}

\newcommand{\subsubsubsection}[1]{\emph{#1.}\xspace}

\newcommand{\n}{$\cdot$}

\newcommand{\code}[1]{\texttt{#1}}

\newcolumntype{L}[1]{>{\raggedright\arraybackslash}m{#1}} % left aligned column

\newboolean{showedits}
\setboolean{showedits}{true} % toggle to show or hide edits
\ifthenelse{\boolean{showedits}}
{
	\newcommand{\del}[1]{\textcolor{red}{\sout{#1}}} % please delete
	\newcommand{\nbe}[3]{
		{\colorbox{#3}{\bfseries\sffamily\scriptsize\textcolor{white}{#1}}}
		{\textcolor{#3}{\sf\small$\blacktriangleright$\textit{#2}$\blacktriangleleft$}}}
}{
	\newcommand{\del}[1]{} % please delete
	
	\newcommand{\nbe}[3]{}
}

\ifthenelse{\boolean{showedits}}
{
	\newtcolorbox{inserted}{
		title=Inserted text:,
		colframe=blue,colback=blue!5!white,
		breakable,
		leftrule=0mm, 
		bottomrule=0mm,
		rightrule=0mm,
		toprule=0mm,
		arc=0mm, outer arc=0mm,
		oversize
	}
	\newtcolorbox{deleted}{
		title=Deleted text:,
		colframe=red,colback=red!5!white,
		breakable,
		leftrule=0mm, 
		bottomrule=0mm,
		rightrule=0mm,
		toprule=0mm,
		arc=0mm, outer arc=0mm,
		oversize
	}
	\newtcolorbox{refactored}{
		title=Rewritten text:,
		colframe=blue,colback=red!5!white,
		breakable,
		leftrule=0mm, 
		bottomrule=0mm,
		rightrule=0mm,
		toprule=0mm,
		arc=0mm, outer arc=0mm,
		oversize
	}
}{

}

\newboolean{showcomments}
\setboolean{showcomments}{true}
\ifthenelse{\boolean{showcomments}}
{\newcommand{\nbc}[3]{
		{\colorbox{#3}{\bfseries\sffamily\scriptsize\textcolor{white}{#1}}}
		{\textcolor{#3}{\sf\small$\blacktriangleright$\textit{#2}$\blacktriangleleft$}}}
	}
{\newcommand{\nbc}[3]{}
	}

\newboolean{isblinded}
\setboolean{isblinded}{false}
\ifthenelse{\boolean{isblinded}}
{\newcommand\blind[1]{BLINDED\xspace}}
{\newcommand\blind[1]{#1\xspace}}

%% file: PhishingOnDemand.bbl
\begin{thebibliography}{10}
\providecommand{\url}[1]{#1}
\csname url@samestyle\endcsname
\providecommand{\newblock}{\relax}
\providecommand{\bibinfo}[2]{#2}
\providecommand{\BIBentrySTDinterwordspacing}{\spaceskip=0pt\relax}
\providecommand{\BIBentryALTinterwordstretchfactor}{4}
\providecommand{\BIBentryALTinterwordspacing}{\spaceskip=\fontdimen2\font plus
\BIBentryALTinterwordstretchfactor\fontdimen3\font minus
  \fontdimen4\font\relax}
\providecommand{\BIBforeignlanguage}[2]{{%
\expandafter\ifx\csname l@#1\endcsname\relax
\typeout{** WARNING: IEEEtran.bst: No hyphenation pattern has been}%
\typeout{** loaded for the language `#1'. Using the pattern for}%
\typeout{** the default language instead.}%
\else
\language=\csname l@#1\endcsname
\fi
#2}}
\providecommand{\BIBdecl}{\relax}
\BIBdecl

\bibitem{Hong:2012}
J.~Hong, ``The state of phishing attacks,'' \emph{Communications of the ACM},
  vol.~55, no.~1, pp. 74--81, 2012.

\bibitem{ProofPoint:2020}
\BIBentryALTinterwordspacing
ProofPoint. (2020) State of the phish, annual report. [Online]. Available:
  \url{https://www.proofpoint.com/sites/default/files/gtd-pfpt-us-tr-state-of-the-phish-2020.pdf}
\BIBentrySTDinterwordspacing

\bibitem{Oest:2018}
A.~Oest, Y.~Safei, A.~Doup{\'e}, G.-J. Ahn, B.~Wardman, and G.~Warner, ``Inside
  a phisher's mind: Understanding the anti-phishing ecosystem through phishing
  kit analysis,'' in \emph{2018 APWG Symposium on Electronic Crime Research
  (eCrime)}.\hskip 1em plus 0.5em minus 0.4em\relax IEEE, 2018, pp. 1--12.

\bibitem{Moore:2007}
T.~Moore and R.~Clayton, ``Examining the impact of website take-down on
  phishing,'' in \emph{Proceedings of the anti-phishing working groups 2nd
  annual eCrime researchers summit}.\hskip 1em plus 0.5em minus 0.4em\relax
  ACM, 2007, pp. 1--13.

\bibitem{Han:2016}
X.~Han, N.~Kheir, and D.~Balzarotti, ``{PhishEye}: Live monitoring of sandboxed
  phishing kits,'' in \emph{Proceedings of the 2016 ACM SIGSAC Conference on
  Computer and Communications Security}, 2016, pp. 1402--1413.

\bibitem{Han:2015}
------, ``The role of cloud services in malicious software: Trends and
  insights,'' in \emph{International Conference on Detection of Intrusions and
  Malware, and Vulnerability Assessment}.\hskip 1em plus 0.5em minus
  0.4em\relax Springer, 2015, pp. 187--204.

\bibitem{McGrath:2008}
D.~K. McGrath and M.~Gupta, ``Behind phishing: An examination of phisher modi
  operandi,'' \emph{LEET}, vol.~8, p.~4, 2008.

\bibitem{Canali:2013}
D.~Canali, D.~Balzarotti, and A.~Francillon, ``The role of web hosting
  providers in detecting compromised websites,'' in \emph{Proceedings of the
  22nd international conference on World Wide Web}, 2013, pp. 177--188.

\bibitem{Cui:2018}
Q.~Cui, G.-V. Jourdan, G.~V. Bochmann, I.-V. Onut, and J.~Flood, ``Phishing
  attacks modifications and evolutions,'' in \emph{European Symposium on
  Research in Computer Security}.\hskip 1em plus 0.5em minus 0.4em\relax
  Springer, 2018, pp. 243--262.

\bibitem{Cova:2008}
M.~Cova, C.~Kruegel, and G.~Vigna, ``There is no free phish: An analysis of
  ``free'' and live phishing kits.'' \emph{WOOT}, vol.~8, pp. 1--8, 2008.

\bibitem{McCalley:2011}
H.~McCalley, B.~Wardman, and G.~Warner, ``Analysis of back-doored phishing
  kits,'' in \emph{IFIP International Conference on Digital Forensics}.\hskip
  1em plus 0.5em minus 0.4em\relax Springer, 2011, pp. 155--168.

\bibitem{Moore:2009}
T.~Moore and R.~Clayton, ``Evil searching: Compromise and recompromise of
  internet hosts for phishing,'' in \emph{International Conference on Financial
  Cryptography and Data Security}.\hskip 1em plus 0.5em minus 0.4em\relax
  Springer, 2009, pp. 256--272.

\bibitem{Birk:2007}
D.~Birk, S.~Gajek, F.~Grobert, and A.-R. Sadeghi, ``Phishing
  phishers---observing and tracing organized cybercrime,'' in \emph{Second
  International Conference on Internet Monitoring and Protection (ICIMP
  2007)}.\hskip 1em plus 0.5em minus 0.4em\relax IEEE, 2007, pp. 3--3.

\bibitem{Hong:2020}
J.~Hong, T.~Kim, J.~Liu, N.~Park, and S.-W. Kim, ``Phishing {URL} detection
  with lexical features and blacklisted domains,'' in \emph{Adaptive Autonomous
  Secure Cyber Systems}.\hskip 1em plus 0.5em minus 0.4em\relax Springer, 2020,
  pp. 253--267.

\bibitem{Jain:2018}
A.~K. Jain and B.~Gupta, ``{PHISH-SAFE}: {URL} features-based phishing
  detection system using machine learning,'' in \emph{Cyber Security}.\hskip
  1em plus 0.5em minus 0.4em\relax Springer, 2018, pp. 467--474.

\bibitem{Rao:2020}
R.~S. Rao, T.~Vaishnavi, and A.~R. Pais, ``{CatchPhish}: detection of phishing
  websites by inspecting {URLs},'' \emph{Journal of Ambient Intelligence and
  Humanized Computing}, vol.~11, no.~2, pp. 813--825, 2020.

\bibitem{Kp:2020}
S.~KP, M.~Alazab \emph{et~al.}, ``Malicious {URL} detection using deep
  learning,'' \emph{TechRxiv}, 2020.

\bibitem{Zamir:2020}
A.~Zamir, H.~U. Khan, T.~Iqbal, N.~Yousaf, F.~Aslam, A.~Anjum, and M.~Hamdani,
  ``Phishing web site detection using diverse machine learning algorithms,''
  \emph{The Electronic Library}, 2020.

\bibitem{Joshi:2009}
Y.~Joshi, D.~Das, and S.~Saha, ``Mitigating man in the middle attack over
  secure sockets layer,'' in \emph{2009 IEEE International Conference on
  Internet Multimedia Services Architecture and Applications (IMSAA)}.\hskip
  1em plus 0.5em minus 0.4em\relax IEEE, 2009, pp. 1--5.

\bibitem{Zhang:2007}
Y.~Zhang, J.~I. Hong, and L.~F. Cranor, ``Cantina: a content-based approach to
  detecting phishing web sites,'' in \emph{Proceedings of the 16th
  international conference on World Wide Web}, 2007, pp. 639--648.

\bibitem{Liu:2006}
W.~Liu, X.~Deng, G.~Huang, and A.~Y. Fu, ``An antiphishing strategy based on
  visual similarity assessment,'' \emph{IEEE Internet Computing}, vol.~10,
  no.~2, pp. 58--65, 2006.

\bibitem{Chiew:2015}
K.~L. Chiew, E.~H. Chang, W.~K. Tiong \emph{et~al.}, ``Utilisation of website
  logo for phishing detection,'' \emph{Computers \& Security}, vol.~54, pp.
  16--26, 2015.

\bibitem{Hara:2009}
M.~Hara, A.~Yamada, and Y.~Miyake, ``Visual similarity-based phishing detection
  without victim site information,'' in \emph{2009 IEEE Symposium on
  Computational Intelligence in Cyber Security}.\hskip 1em plus 0.5em minus
  0.4em\relax IEEE, 2009, pp. 30--36.

\bibitem{Bicakci:2014}
K.~Bicakci, D.~Unal, N.~Ascioglu, and O.~Adalier, ``Mobile authentication
  secure against man-in-the-middle attacks,'' in \emph{2014 2nd IEEE
  International Conference on Mobile Cloud Computing, Services, and
  Engineering}.\hskip 1em plus 0.5em minus 0.4em\relax IEEE, 2014, pp.
  273--276.

\bibitem{Ulqinaku:2019}
E.~Ulqinaku, D.~Lain, and S.~Capkun, ``{2FA-PP}: 2nd factor phishing
  prevention,'' in \emph{Proceedings of the 12th Conference on Security and
  Privacy in Wireless and Mobile Networks}, 2019, pp. 60--70.

\end{thebibliography}
